\documentclass[11pt,a4paper]{article}
\usepackage{jheppub,amsmath, amsthm, amssymb,slashed,url}
\usepackage{graphicx}
\usepackage{epstopdf}

\def\OMIT#1{{}}

\newcommand{\bit}[1]{\mbox{\boldmath$#1$}}
\newcommand{\ft}[2]{{\textstyle\frac{#1}{#2}}}

\newcommand{\nn}{\nonumber}

\def\OMIT#1{{}}
\def\nc{{N_c}}

\def\nc{{N_c}}

\def\yo2{{f_\pi^2}}
\def\llra{{\relbar\joinrel\longrightarrow}}
\def\mapright#1{{\smash{\mathop{\llra}\limits_{#1}}}}

\def\oneht{\textstyle{1\over 2} }

\newcommand{\BE}{\begin{equation}}
\newcommand{\EE}{\end{equation}}
\newcommand{\BA}{\begin{eqnarray}}
\newcommand{\EA}{\end{eqnarray}}

\font\teneurm=eurm10 \font\seveneurm=eurm7 \font\fiveeurm=eurm5
\newfam\eurmfam
\textfont\eurmfam=\teneurm \scriptfont\eurmfam=\seveneurm
\scriptscriptfont\eurmfam=\fiveeurm

\font\teneusm=eusm10 \font\seveneusm=eusm7 \font\fiveeusm=eusm5
\newfam\eusmfam
\textfont\eusmfam=\teneusm \scriptfont\eusmfam=\seveneusm
\scriptscriptfont\eusmfam=\fiveeusm

\font\tencmmib=cmmib10 \skewchar\tencmmib='177
\font\sevencmmib=cmmib7 \skewchar\sevencmmib='177
\font\fivecmmib=cmmib5 \skewchar\fivecmmib='177
\newfam\cmmibfam
\textfont\cmmibfam=\tencmmib \scriptfont\cmmibfam=\sevencmmib
\scriptscriptfont\cmmibfam=\fivecmmib

\def\Pi{\varPi}

\title{Broken Chiral Symmetry on a Null Plane}

 \author{Silas R.~Beane\footnote{Address as of 1 September, 2013: Department of Physics, University of Washington.}}

\affiliation{Helmholtz-Institut f\"ur Strahlen- und Kernphysik (Theorie),
\\ Universit\"at Bonn, D-53115 Bonn, Germany
\\beane@hiskp.uni-bonn.de\ \ +49 228 733452}

\date{\mydate}

\abstract{On a null-plane (light-front), all effects of spontaneous
  chiral symmetry breaking are contained in the three Hamiltonians
  (dynamical Poincar\'e generators), while the vacuum state is a
  chiral invariant. This property is used to give a general proof of
  Goldstone's theorem on a null-plane. Focusing on null-plane QCD with
  $N$ degenerate flavors of light quarks, the chiral-symmetry
  breaking Hamiltonians are obtained, and the role of vacuum
  condensates is clarified.  In particular, the null-plane
  Gell-Mann-Oakes-Renner formula is derived, and a general
  prescription is given for mapping all chiral-symmetry breaking QCD
  condensates to chiral-symmetry conserving null-plane QCD
  condensates. The utility of the null-plane description lies in the
  operator algebra that mixes the null-plane Hamiltonians and the
  chiral symmetry charges. It is demonstrated that in a certain
  non-trivial limit, the null-plane operator algebra reduces to the
  symmetry group $SU(2N)$ of the constituent quark model.}
\begin{document} \maketitle

\section{Introduction}
\label{intro}

\noindent Spontaneous symmetry breaking is usually treated as a
phenomenon that arises from properties of an asymmetric quantum mechanical vacuum
state. In particular, the non-invariance of the vacuum
state with respect to a symmetry is said to lead to spontaneous
symmetry breakdown. While this picture is clearly valid and useful, it
is not generally appreciated that in relativistic theories of quantum mechanics, 
it is strictly a matter of convention which
arises from the (usually implicit) choice of quantization
surface~\cite{Dirac:1949cp}. Indeed, the standard viewpoint ---the
instant form--- arises from choosing to view dynamics in Minkowski
space as the evolution of families of parallel spaces at various
instants of time. An alternate view of dynamics is to consider the
evolution of families of parallel spaces tangent to the light cone;
i.e.~null planes~\cite{Dirac:1949cp,Weinberg:1966jm,Susskind:1967rg,Bardakci:1969dv,Kogut:1969xa,Leutwyler:1977vy}. In this
viewpoint ---the front form--- the momentum operator has a spectrum
confined to the open positive half-line and therefore the vacuum of
the interacting theory may be regarded as the structureless Fock-space
vacuum, which is an invariant with respect to all internal symmetries,
and spontaneous symmetry breaking must be attributed to properties of
the dynamical Poincar\'e generators. Therefore in the front form,
spontaneous chiral symmetry breaking is a property of operators rather
than of a complicated vacuum state.  Naturally one expects that
physics is independent of the choice of quantization surface. However,
for theories like QCD where the detailed dynamics are largely
intractable, one may suppose that the two forms of dynamics lead to
distinct insights into the behavior of the theory at strong coupling.
Our goal in this paper is to argue that this is indeed the case.

The fundamental point which we wish to emphasize in this paper is
that, in contrast to the instant form, where spontaneous symmetry
breaking lives entirely in the non-trivial vacuum, in the front form,
symmetry breaking is expressed entirely through the fact that the
Hamiltonians, or dynamical Poincar\'e generators, do not commute with
the internal symmetry charges. The resulting commutation relations
among space-time generators and internal symmetry generators in QCD
imply powerful constraints on the spectrum and spin of the hadronic
world~\cite{Weinberg:1969hw,Weinberg:1969db,Weinberg:1990xn,Weinberg:1994tu}.
There have been many studies of spontaneous chiral symmetry breaking
on null
planes~\cite{Jersak:1969zg,Leutwyler:1969av,Weinberg:1969hw,Weinberg:1969db,Feinberg:1973qb,Eichten:1973ip,Casher:1973vh,Casher:1974xd,Carlitz:1974sg,Sazdjian:1974gk,Wilson:1994fk,Wilson:1994gn,Susskind:1994wr,Kim:1994rm,Burkardt:1995eb,Burkardt:1996pa,Burkardt:1998dd,Yamawaki:1998cy,Itakura:2001yt,Burkardt:2002yf,Wu:2003vn,Lenz:2004tw,Dalley:2004re,Brodsky:2008xm,Brodsky:2008xu,Brodsky:2009zd,Ji:2009jc,Strikman:2010pu,Chang:2011mu,Alberg:2012wr,Brodsky:2010xf,Schweitzer:2012hh,Brodsky:2012ku}.
In many cases the emphasis has been on learning detailed information
about the dynamical mechanism of chiral symmetry breaking in QCD and
in models. Here our approach is much less ambitious; we assume that
chiral symmetry is broken spontaneously by complicated and not-well
understood dynamics, and we then determine the constraints that follow
from this assumption. In particular, we are interested primarily in
formulating the model-independent consequences of chiral symmetry
breaking on null-planes. A fundamental assumption we make is that
physics must be independent of the choice of quantization
surface. Nowhere in this study do we find anything resembling a
contradiction of this basic assumption. Indeed, this assumption of
what one might call ``form invariance'' leads to various constraints
which reveal a great deal about the nature and consistency of chiral
symmetry breaking on null-planes. On general grounds, the null-plane
chiral symmetry charges annihilate the vacuum. Therefore, in order
that spontaneous chiral symmetry breaking take place, the chiral
symmetry axial-vector current on the null-plane cannot be
conserved~\cite{Jersak:1969zg,Leutwyler:1969av}.  This property leads
to a simple proof of Goldstone's theorem on a null-plane, which is
completely decoupled from any assumptions about the formation of
symmetry-breaking condensates. A second consistency condition is that
the part of the QCD vacuum energy that is dependent on the quark
masses should be invariant with respect to the choice of
coordinates. This condition recovers the Gell-Mann-Oakes-Renner
relation~\cite{GellMann:1968rz} in the null-plane description, and
leads to a general prescription for relating all instant-form
chiral-symmetry breaking condensates to the vacuum expectation values
of chiral singlet null-plane QCD operators.

It is difficult to find a general solution of the null-plane operator
algebra~\cite{Weinberg:1969hw,Weinberg:1969db,Weinberg:1990xn,Weinberg:1994tu}. However,
there is a non-trivial limit in which a solution can be found. One
expects that, in general, the chiral symmetry breaking part of the
null-plane energy has an energy scale comparable to $\Lambda_{QCD}$
and therefore is not parametrically small. However, assuming that this
is small (which is the case parametrically for baryon operators at
large-$N_c$), while the chiral symmetry breaking part of the spin
Hamiltonians is of natural size, allows a non-trivial solution of the
operator algebra which closes to the Lie brackets of $SU(2N)$, thus
recovering the basic group theoretical structure of the constituent
quark model. This result, originally found by
Weinberg~\cite{Weinberg:1994tu} working with current-algebra sum rules
in special Lorentz frames, is shown in this context to be a general
consequence of the null-plane QCD Lie algebraic constraints which are
valid in any Lorentz frame.

The paper is organized as follows.  In section~\ref{sec:poincare}, the
null-plane coordinates and conventions are introduced, and the
front-form Poincar\'e algebra is obtained. The null-plane Hamiltonians
and the Lie brackets that they satisfy are identified, and the
momentum eigenstates are constructed.  In
section~\ref{sec:chiralityintro} the null-plane internal symmetry
charges are introduced, and the commutators that mix Poincar\'e and
chiral generators are obtained. Using these commutators, a general
proof of Goldstone's theorem is given, and a polology analysis is
given which elucidates the structure of the axial-vector current on
the null-plane.  The special case of QCD with $N$ flavors of light
quarks is considered in section~\ref{sec:qcd}. The QCD Lagrangian is
expressed in the null-plane coordinates, and the chiral symmetry
breaking Hamiltonians and the constraints that they satisfy are
derived.  The issue of condensates in the null-plane formulation is
addressed in detail; the Gell-Mann-Oakes-Renner formula is recovered
in the front-form and a general method for relating instant-form
condensates to front-form condensates is presented.
Section~\ref{sec:consequences} explores the consequences of the QCD
null-plane operator algebra. In particular, a simple solution of the
operator algebra is given which contains the spin-flavor symmetries of
the constituent quark model.  In section~\ref{sec:conc} we summarize
our findings and conclude.  \vskip0.3in
\noindent {\it Nota bene}: 
We have made use of the many general
reviews of null-plane (or light-front)
quantization~\cite{Hornbostel:1990ya,Perry:1994kp,Zhang:1994ti,Burkardt:1995ct,Harindranath:1996hq,Brodsky:1997de,Perry:1997uv,Miller:1997cr,Venugopalan:1998zd,Heinzl:2000ht,Miller:2000kv,Diehl:2003ny,Belitsky:2005qn}, as well as reviews that focus
primarily on chiral symmetry related
issues~\cite{Mustaki:1994mf,Yamawaki:1998cy,Itakura:2001yt}.
In order to provide a self-contained description of the subject
of chiral-symmetry breaking on a null-plane, there is a significant 
amount of review material in this paper.

%
\section{Space-time symmetry in the front form}
\label{sec:poincare}

\subsection{A null plane defined}

\begin{figure}[!t]
  \centering
     \includegraphics[scale=0.54]{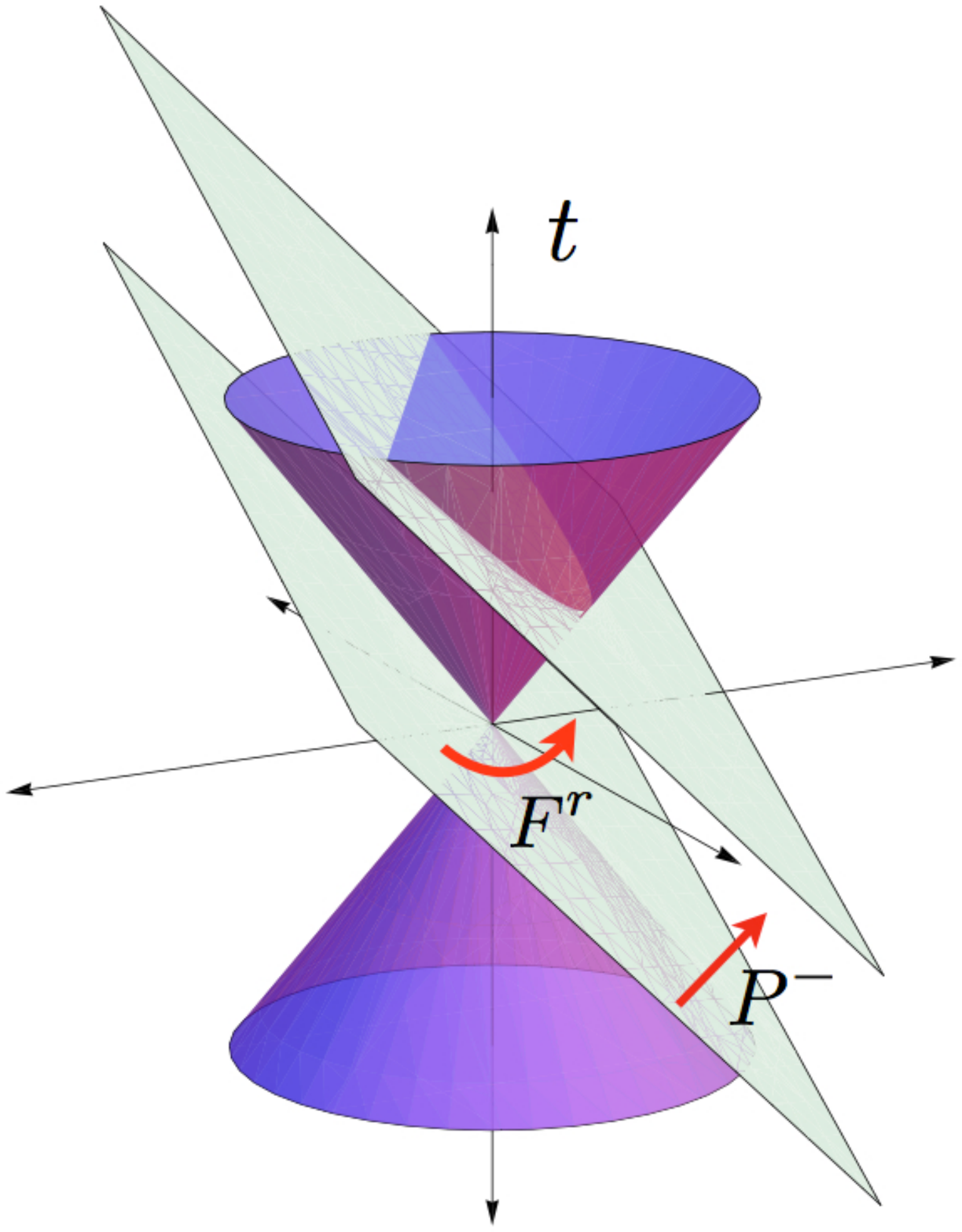}
     \caption{A null plane is a surface tangent to the light cone. The
       null-plane Hamiltonians map the initial light-like surface onto
       some other surface and therefore describe the dynamical
       evolution of the system. The energy $P^-$ translates the system
       in the null-plane time coordinate $x^+$, whereas the spin
       Hamiltonians $F_r$ rotate the initial surface about the surface
       of the light cone.}
  \label{fig:nullplane}
\end{figure}

\noindent In the front-form of relativistic Hamiltonian dynamics, one
chooses the initial state of the system to be on a light-like plane,
or null-plane, which is a hypersurface of points $x$ in Minkowski
space such that $x\cdot n =\tau$ (see fig.~\ref{fig:nullplane}). Here
$n$ is a light-like vector which will be chosen below, and $\tau$ is a
constant which plays the role of time. We will refer to a null-plane
as $\Sigma_n^\tau$.  The subgroup of the Poincar\'e group that maps
$\Sigma_n^\tau$ to itself is called the stability group of the
null-plane and determines the kinematics within the null-plane. The
remaining three Poincar\'e generators map $\Sigma_n^\tau$ to a new
surface, $\Sigma_n^{\tau'}$, and therefore describe the evolution of the
system in time. The front-form is special in that it has seven
kinematical generators, the largest stability group of all of the
forms of dynamics~\cite{Dirac:1949cp}. It stands to reason that in
complicated problems in relativistic quantum mechanics one would
prefer a formulation which has the fewest number of Hamiltonians to
determine.
\subsection{Choice of coordinates}

\noindent Consider the light-like vectors $n^\mu$ and $n^{\ast\mu}$
which satisfy $n^2 = n^{\ast 2} = 0$ and $n \cdot n^\ast = 1$. Here we will
choose these vectors such that 
\begin{equation}
n^\mu \equiv \ft{1}{\sqrt{2}} (1, 0, 0, -1) \quad , \quad
n^{\ast \mu} \equiv \ft{1}{\sqrt{2}} (1, 0, 0, 1)
\, .
\label{nullvectors}
\end{equation}
We will take the initial surface to be the null-plane
$\Sigma_n^0$. A coordinate system adapted to null-planes is then given by
\BA
x^+ \equiv x \cdot n = \ft{1}{\sqrt{2}} (x^0 + x^3)
\, , \qquad
x^- \equiv x \cdot n^\ast = \ft{1}{\sqrt{2}} (x^0 - x^3)
\label{eq:LCcoordinates}
\EA which we take as the time variable and ``longitudinal'' position, respectively~\footnote{This
  is known as the Kogut-Soper convention~\cite{Kogut:1969xa}. Our
  metric and other notational conventions can be found in
  Appendix~\ref{npconventions} and in Ref.~\cite{Brodsky:1997de}.}.  The remaining coordinates, ${\bf
  x}_\perp=( {x}^1, {x}^2)$ provide the ``transverse''
position.  Denoting the null-plane contravariant coordinate
four-vector by ${\tilde x}^\mu=(x^+,{x}^1,{x}^2,x^-)=(x^+,{\bf x}_\perp,x^-)$, then
one can write
\BA
{\tilde x}^\mu\ = \ {\cal C}^\mu_\nu\ x^\nu \ .
\label{eq:ETtoLC}
\EA The matrix ${\cal C}^\mu_\nu$, given explicitly in
Appendix~\ref{npconventions}, allows one to transform all Lorentz
tensors from instant-form to front-form coordinates. In particular,
the null-plane metric tensor is given by
\BA
{\tilde g}_{\mu\nu}\ =\ ({\cal C}^{-1})^\alpha_\mu\ g_{\alpha\beta}\ ({\cal C}^{-1})^\beta_\nu \ .
\label{eq:LCmetric1}
\EA

The energy, canonical to the null-plane time variable $x^+$ is
$p^-=p_+$ , and the momentum canonical to the longitudinal position
variable $x^-$ is $p^+=p_-$.  Therefore, the on-mass-shell condition
for a relativistic particle of mass $m$ yields the null-plane
dispersion relation:
\BA
p^-\ =\ \frac{{\bf p}^2_\perp\ +\ m^2}{2p^+} \ .
\label{eq:LCdisp}
\EA This dispersion relation reveals several interesting generic
features of the null-plane formulation. Firstly, the dispersion
relation resembles the non-relativistic dispersion relation of a
particle of mass $p^+$ in a constant potential. Secondly, we see that
the positivity and finiteness of the null-plane energy of a free
massive particle requires $p^+>0$. Only massless particles with
strictly vanishing momentum can have $p^+=0$. This implies that pair
production is subtle, and the vacuum state is in some sense simple,
with the exception of contributions that are strictly from $p^+=0$
modes~\cite{Hornbostel:1990ya,Perry:1994kp,Zhang:1994ti,Burkardt:1995ct,Harindranath:1996hq,Brodsky:1997de,Perry:1997uv,Venugopalan:1998zd,Heinzl:2000ht,
  Diehl:2003ny,Belitsky:2005qn}.

%
\subsection{The null-plane Poincar\'e generators}

\noindent In this section we will review the
Lie brackets of the Lorentz generators in the front form~\footnote{Here
  we follow closely the development of
  Refs.~\cite{Bardakci:1969dv,Kogut:1969xa,Leutwyler:1977vy}. See also Ref.~\cite{Ballesteros:1995mi}.}.  The Poincar\'e algebra
in our convention is:
\BA
& & \lbrack\, P^\mu \, ,\, P^\nu\, \rbrack\ =\ 0 \qquad , \qquad \lbrack\, { M}_{\mu\nu} \, ,\, P_\rho\, \rbrack\ 
=\ i\, (\,  g_{\nu\rho} P_\mu\, -\, g_{\mu\rho} P_\nu \, ) 
\nonumber \\
& &\lbrack\, { M}_{\mu\nu} \, ,\, { M}_{\rho\sigma}\, \rbrack\  = 
\ i\, (\, g_{\mu\sigma} { M}_{\nu\rho}\, + \, g_{\nu\rho} { M}_{\mu\sigma}\, - \, g_{\mu\rho} { M}_{\nu\sigma}\, -
\, g_{\nu\sigma} { M}_{\mu\rho}\, ) \ ,
\label{eq:ETpoincare}
\EA where ${ M}_{ij}=\epsilon_{ijk}{ J}_k$ and ${
  M}_{i0}={ K}_i$ with ${ J}_i$ and ${ K}_i$ the generators
of rotations and boosts, respectively.  Using ${\cal C}^\mu_\nu$ we
can transform from the instant-form to the front-form giving ${\tilde
  { P}}^\mu=({ P}^+,{ P}^1,{ P}^2,{ P}^-)$, 
${\tilde {  M}}_{r+}=-{\tilde {  M}}_{+r}={F}_r$, 
${\tilde {  M}}_{r-}=-{\tilde {  M}}_{-r}={E}_r$, 
${\tilde { M}}_{rs}=\epsilon_{rs}{J}_3$, and
${\tilde { M}}_{+-}=-{\tilde { M}}_{-+}={ K}_3$, where we have
defined
\BA
{ P}^+ &=& \textstyle{1\over\sqrt{2}}\, (\, { P}^0\, +\, { P}^3\, )\qquad , \qquad { P}^-\ =\ \textstyle{1\over\sqrt{2}}\, (\, { P}^0\, -\, { P}^3\, )\ ;
\nonumber \\
{ E}_r &=& \textstyle{1\over\sqrt{2}}\, (\, { K}_r\, + \, \epsilon_{rs}{ J}_s\, )\quad , \quad 
\quad{ F}_r\ =\ \textstyle{1\over\sqrt{2}}\, (\, { K}_r\, - \, \epsilon_{rs}{ J}_s\, )\ .
\label{eq:LCpoingens}
\EA Here ${ P}_+={ P}^-$ is the null-plane energy while ${ P}_- = { P}^+$ is the longitudinal momentum.
(Note that the indices $r,s,t,\ldots$ are transverse indices that range over $1,2$.
See Appendix~\ref{npconventions}.)

It is straightforward to show that ${ P}^+$, ${ P}_r$, ${ K}_3$, ${ E}_r$,
and ${ J}_3$ are kinematical generators that leave the null plane $x^+=0$
intact. These seven generators form the stability group of the null plane. 
It is useful to classify the subgroups of the Poincar\'e algebra by
considering the transformation properties of the generators with respect
to longitudinal boosts, which serve to rescale the generators. Writing
\BA
& & \lbrack\, { K}_3 \, ,\, A\, \rbrack\ =\ -i \gamma\, A
\label{eq:goodness}
\EA
where $A$ is a generator, one finds ${ E}_r$ and ${ P}^+$ have $\gamma=1$,
${ J}_3$, ${ K}_3$ and ${ P}_r$ have $\gamma=0$, and ${ P}^-$ and ${ F}_r$ have $\gamma=-1$.
The Poincar\'e generators have subgroups $G_\gamma$ labeled by $\gamma$, and
there exist two seven-parameter subgroups $S_\pm$ with a semi-direct product
structure $S_\pm=G_0\times G_\pm$. Therefore the stability group coincides
with the subgroup $S_+$.
The non-vanishing commutation relations among these generators are:
\BA
& & \lbrack\, { K}_3 \, ,\, { E}_r\, \rbrack\ =\ -i { E}_r \quad , \quad \lbrack\, { K}_3 \, ,\, { P}^+\, \rbrack\ =\ -i { P}^+  \ ;
\nonumber \\
& & \lbrack\, { J}_3 \, ,\, { E}_r\, \rbrack\ =\ i\epsilon_{rs} { E}_s \quad , \quad 
\lbrack\, { J}_3 \, ,\, { P}_r\, \rbrack\ =\ i\epsilon_{rs} { P}_s \ ;
\nonumber \\
& & \qquad \qquad \quad
\lbrack\, { E}_r  \, ,\,{ P}_s \, \rbrack\ =\ -i\delta_{rs} { P}^+ \ .
\label{eq:LCSGcomms}
\EA

By contrast, ${ P}^-$ and ${ F}_r$ are the Hamiltonians which consist
of the subgroup $G_{-1}$; they are
the dynamical generators which move physical states away from the
$x^+=0$ surface (see fig.~\ref{fig:nullplane}). The non-vanishing commutators among the stability
group generators and the Hamiltonians are:
\BA
& & \lbrack\, { K}_3 \, ,\, { P}^-\, \rbrack\ =\ i { P}^- \quad , \quad \lbrack\, { E}_r \, ,\, { P}^-\, \rbrack\ =\ -i { P}_r \ ;
\nonumber \\
& & \lbrack\, { K}_3 \, ,\, { F}_r\, \rbrack\ =\ i { F}_r \quad , \quad \lbrack\, { J}_3 \, ,\, { F}_r\, \rbrack\ =\ i\epsilon_{rs} { F}_s \ ;
\nonumber \\
& & \lbrack\, { P}_r \, ,\, { F}_s\, \rbrack\ =\ i\delta_{rs}{ P}^- \quad , \quad \lbrack\, { P}^+ \, ,\, { F}_r\, \rbrack\ =\ i{ P}_r\ ;
\nonumber \\
& & \lbrack\, { E}_r \, ,\, { F}_s\, \rbrack\ =\ -i\left(\,\delta_{rs}\,{ K}_3\,+\,\epsilon_{rs} { J}_3\,\right)\ .
\label{eq:LCHSGcomms}
\EA 
This algebraic structure is isomorphic to the Galilean group of
two-dimensional quantum mechanics where one identifies
$\lbrace\,P^-,E_r,P_r,J_3,P^+\,\rbrace$ with the Hamiltonian, Galilean
boosts, momentum, angular momentum, and mass, respectively. This
isomorphism is responsible for the similarities between the front form
and nonrelativistic quantum mechanics that we noted in the dispersion
relation, and was originally noted in the context of the
infinite momentum frame of instant-form
dynamics~\cite{Weinberg:1966jm,Susskind:1967rg} which has a similar
dispersion relation.

%
\subsection{Null-plane momentum states and reduced Hamiltonians}

\noindent As momentum is a kinematical observable, it is convenient to work with momentum eigenstates,
such that
\BA
P_r\,|\, p^+\, ,\, {\bf p}_\perp\,\rangle  &=& p_r|\, p^+\, ,\, {\bf p}_\perp\,\rangle \ ;\\
P^+\,|\, p^+\, ,\, {\bf p}_\perp\,\rangle  &=& p^+|\, p^+\, ,\, {\bf p}_\perp\,\rangle \ .
\label{eq:momonmom}
\EA
The action of the boosts on momentum states follows directly from the commutation relations in eq.~\ref{eq:LCSGcomms}
and is given by
\BA
e^{-iv_r E_r}e^{-i\omega K_3}|\, p^+\, ,\, {\bf p}_\perp\, \rangle = |\, e^\omega\,p^+\, ,\, {\bf p}_\perp+p^+{\bf v}_\perp\,\rangle \ .
\label{eq:boostsdefined}
\EA
One can then define the unitary boost operator
\BA
{\cal U}(p^+,p_r)=e^{-i\beta_r E_r}e^{-i\beta_3 K_3} \ ,
\label{eq:boostop}
\EA
with $\beta_r\equiv p_r/p^+$ and $\beta_3\equiv\log(\sqrt{2}p^+/M)$ 
which boosts the state at rest to one with arbitrary momentum:
\BA
{\cal U}(p^+,p_r)|\, M/\sqrt{2}\, ,\, {\bf 0}\, \rangle = |\, p^+\, ,\, {\bf p}_\perp\,\rangle \ .
\label{eq:boostfromrest}
\EA
The action of the boosts on the momentum states is then easily found to be
\BA
E_r\,|\, p^+\, ,\, {\bf p}_\perp\,\rangle  &=& ip^+\frac{d}{dp_r}|\, p^+\, ,\, {\bf p}_\perp\,\rangle \ ;\\
K_3\,|\, p^+\, ,\, {\bf p}_\perp\,\rangle  &=& ip^+\frac{d}{dp^+}|\, p^+\, ,\, {\bf p}_\perp\,\rangle \ .
\label{eq:boostonmom}
\EA

Unitarity of the boost operators fixes the normalization of the momentum states up to a constant. We assume the covariant
normalization:
\BA
\langle\, p^{+{\prime}}\, ,\, {\bf p}_\perp^{\,\prime}\, | \, p^+\, ,\, {\bf p}_\perp\, \rangle\  =\ 
(2\pi)^3\,2\,p^+\,
\delta(\,p^{+{\prime}}\,-\,p^+\,)\,\delta^2(\,{\bf p}_\perp^{\,\prime}\,-\,{\bf p}_\perp \,) \ ,
\label{eq:HBnormalization}
\EA
and the corresponding completeness relation
\BA
{\bf 1}\, =\, 
\int\,\frac{dp^+d^2{\bf p}_\perp}{(2\pi)^3 2p^+}\,
|\, p^+\, ,\, {\bf p}_\perp\, \rangle\, \langle\, p^+\, ,\, {\bf p}_\perp\, | \ .
\label{eq:HBcomplete}
\EA

We can now find angular momentum operators, ${\cal J}_r$ and ${\cal J}_3$, that are
valid in any frame by boosting from an arbitrary momentum state to a state at rest,
acting with the angular momentum generators $J_r=\epsilon_{rs}(F_s-E_s)/\sqrt{2}$ and $J_3$, and then boosting back
to the arbitrary momentum state. That is,
\BA
{\cal J}_i|\, p^+\, ,\, {\bf p}_\perp\,\rangle \ =\ {\cal U}(p^+,p_r)\;J_i\;{\cal U}^{-1}(p^+,p_r)\,|\, p^+\, ,\, {\bf p}_\perp\,\rangle \ .
\label{eq:generalAM}
\EA
Using this procedure one finds angular momentum operators that are valid in any frame:
\BA
{\cal J}_3 &=& { J}_3\ +\ \epsilon_{rs}\,{ E}_r{ P}_s\,\left(1/{ P}^+\right) \ ; \label{eq:npJ3generalA} \\
{\cal J}_r &=& \epsilon_{rs}\,\big\lbrack\,{ P}^+{ F}_s\ -\ { P}^-{ E}_s\ +\ \epsilon_{st}{ P}_t{\cal J}_3\ +\ { P}_s{ K}_3\,\big\rbrack\,\left(1/M\right) \ .
\label{eq:npJ3generalB}
\EA
Inverting eq.~\ref{eq:npJ3generalB} one then finds the following expressions for the Hamiltonians:
\BA
{ P}^- &=& \left(1/2{ P}^+\right)\,\big\lbrack\, { P}_1^2\ +\ { P}_2^2\ +\ M^2 \,\big\rbrack \ ; \nonumber \\
{ F_1}\ &=& \left(1/{ P}^+\right)\,\big\lbrack\,-{ P}_1{ K}_3\ +\ { P}^-{ E}_1\ -\ { P}_2{\cal J}_3\ -\ M{\cal J}_2\,\big\rbrack  
\ ; \nonumber \\
{ F}_2\ &=& \left(1/{ P}^+\right)\,\big\lbrack\,-{ P}_2{ K}_3\ +\ { P}^-{ E}_2\ +\ { P}_1{\cal J}_3\ +\ M{\cal J}_1\, \big\rbrack \ .
\label{eq:npHams}
\EA 
A striking feature of the null-plane formulation is that the
fundamental dynamical objects are the {\it products} $M^2$ and $M{\cal
  J}_{r}$, rather than the generators themselves. Following
Ref.~\cite{Leutwyler:1977vy}, we will refer to these objects as
reduced Hamiltonians. The reduced Hamiltonians, together with 
${\cal J}_3$, commute with all kinematical generators and
satisfy the algebra of $U(2)$. This is conveniently demonstrated
by making use of the Pauli-Lubanski vector
\BA
{ W}^\mu = \ft12 \varepsilon^{\mu\nu\rho\sigma} { P}_\nu { M}_{\rho\sigma} \ ,
\EA
which satisfies $W^\mu P_\mu=0$ and the non-trivial commutation relations:
\BA
\lbrack\, { M}_{\mu\nu} \, ,\, W_\rho\, \rbrack & =& i\, (\,  g_{\nu\rho} W_\mu\, -\, g_{\mu\rho} W_\nu \, ) \ ;\\
\lbrack\, W^\mu \, ,\,W^\nu\, \rbrack &=& -i\varepsilon^{\mu\nu\rho\sigma} W_\rho P_\sigma \ .
\label{eq:ETpoincareW}
\EA
One then finds general, compact expressions for the angular momentum operators:
\BA
{\cal J}_3 \ = \ { W}^+/{ P}^+ \quad , \quad
M{\cal J}_r \ = \ { W}_r\ -\ { P}_r\;{ W}^+/{ P}^+ \ .
\EA
By considering the commutation relations among $W_\mu$, $P^\mu$ and $M_{\mu\nu}$
one confirms that
\BA
[\, {\cal J}_3 \, ,\,  M{\cal J}_r  \, ] \ =\   i\,\epsilon_{rs} M{\cal J}_s
\ \  \ & , &\  \ \ 
[\, {\cal J}_3 \, ,\,  M^2  \, ]\, = \, 0 \ ; \nonumber \\
{} [\, M{\cal J}_r \, ,\,  M{\cal J}_s  \, ] \ =\  i\,\epsilon_{rs} M^2 {\cal J}_3  
\ \ \ & ,& \ \ \
[\, M^2 \, ,\,  M{\cal J}_r  \, ]\, = \, 0 \ .
\label{eq:dynalgarbB}
\EA 
Hence, the reduced Hamiltonians together with the stability group
generator ${\cal J}_3$ satisfy the algebra of $U(2)$, and the
problem of finding a Lorentz invariant description of a relativistic
quantum mechanical system is thus equivalent to finding a representation
of the three reduced Hamiltonians which satisfy this algebra~\footnote{Since 
the mass operator, $M=\sqrt{p_\mu p^\mu}$, commutes with the spin operators, this
algebra can clearly be expressed in the canonical form: 
$[\, {\cal J}_i \, ,\,  {\cal J}_j  \, ]\, = \, i\epsilon_{ijk}J_k$ and $[\, M \, ,\,  {\cal J}_i  \, ]\, = \, 0$.}.  
Since the essence of Lorentz invariance resides in these Lie brackets, and they
require knowledge of the reduced Hamiltonians, in theories with complicated
dynamics like QCD, the formulation of the theory at weak coupling ---where QCD is
defined as a continuum quantum field theory--- will
lack manifest Lorentz invariance, which is tied up with the detailed
dynamics of the theory, and is as complicated to achieve as finding
the spectrum of the theory.

We can write a general momentum eigenstate as:
\BA
|\, p^+\, ,\, {\bf p}_\perp\, ; \lambda\, ,\,n\, \rangle = |\, p^+\, ,\, {\bf p}_\perp\,\rangle\otimes | \lambda \, ,\, n\, \rangle \ .
\label{eq:HBdefined}
\EA
Here $n$ are additional variables that may be needed to specify the
state of a system at rest, and $\lambda$ is helicity, the eigenvalue of ${\cal J}_3$:
\BA
{\cal J}_3\;|\, p^+\, ,\, {\bf p}_\perp\, ; \lambda\, ,\,n\, \rangle  \ =\ \lambda\;|\, p^+\, ,\, {\bf p}_\perp\, ; \lambda\, ,\,n\, \rangle \ ,
\label{eq:J3eigen}
\EA 
and therefore, using eq.~\ref{eq:npJ3generalA}, we have
\BA
J_3\;|\, p^+\, ,\, {\bf p}_\perp\, ; \lambda\, ,\,n\, \rangle  
\ =\ \left(\,\lambda+i\epsilon_{rs}\,p_r\frac{d}{dp_s}\,\right)\;|\, p^+\, ,\, {\bf p}_\perp\, ; \lambda\, ,\,n\, \rangle \ ,
\label{eq:J3atrest}
\EA 
which completes the catalog of the action of the stability group generators on the momentum states.
It is useful to write
\BA
|\, p^+\, ,\, {\bf p}_\perp\, ; \lambda\, ,\,n\, \rangle \ =\  {\cal U}(p^+,p_r)|\, M/\sqrt{2}\, ,\, {\bf 0}\, ; \lambda\, ,\,n\,\rangle
\ \equiv \ a_n^\dagger\left(p^+\, ,\, {\bf p}_\perp\, ; \lambda\right)\;|\, 0\,\rangle \ ,
\label{eq:vacuumdefined}
\EA
where $a_n^\dagger$ is an operator that creates the momentum state when acting on the null-plane vacuum, $|\, 0\,\rangle$.
What is special about the null-plane description is that the
kinematical generators (with the exception of ${\cal J}_3$) act on
states in a manner independent of the inner variables $n$. And the
reduced Hamiltonians act exclusively on the inner variables in a
manner independent of the momentum. Therefore, one may view the
Poincar\'e algebra by the direct sum of ${\cal K}$ and ${\cal D}$, where
${\cal K}=\lbrace\,E_r,P_r,K_3,P^+\,\rbrace$ contains all stability
group generators with the exception of ${\cal J}_3$ which is grouped
with the reduced Hamiltonians, ${\cal D}=\lbrace\,{\cal J}_3,M{\cal
  J}_r,M^2\rbrace$~\cite{Leutwyler:1977vz}.  

The structure of the Poincar\'e algebra in the front-form is well
suited to the study of systems with complicated dynamics like QCD, as
the dynamical generators are directly related to the most important
observable quantities, namely the energy and the angular momentum of
the system, while momenta and boosts are purely kinematical and
therefore are easy to implement~\footnote{By contrast, in the instant form
  of dynamics, the energy and the boosts are dynamical. As boosts are
  not among the observables, one refers only to the one Hamiltonian
  corresponding to energy.}. The reduced Hamiltonians will have a
fundamental role to play in the description of chiral symmetry
breaking on null planes.

%
\section{Chiral symmetry in the front form}
\label{sec:chiralityintro}

\subsection{Null plane charges and the chiral algebra}

\noindent Consider a Lagrangian field theory that has an
$SU(N)_R\otimes SU(N)_L$ chiral symmetry.  Let us assume that this
system has a null-plane Lagrangian formulation which allows one, by
the standard Noether procedure, to obtain the currents ${\tilde
  J}_\alpha^\mu(x)$ and ${\tilde J}_{5\alpha}^\mu(x)$, which are
related to the symmetry currents via ${\tilde J}^\mu_{L\alpha} =
({\tilde J}^\mu_\alpha - {\tilde J}^\mu_{5\alpha})/2$ and ${\tilde
  J}^\mu_{R\alpha} = ({\tilde J}^\mu_\alpha + {\tilde
  J}^\mu_{5\alpha})/2$.  We will further assume that the Lagrangian
contains an operator that explicitly breaks the chiral symmetry in the
pattern $SU(N)_R\otimes SU(N)_L\rightarrow SU(N)_F$ and is governed by
the parameter $\epsilon_\chi$ such that as $\epsilon_\chi\rightarrow
0$, the symmetry is restored at the classical level. The general
relation between currents and their associated charges is given by
\BA
{Q}(n\cdot x)\, =\, \int\, d^4y\, \delta(n\cdot (x\, -\, y\,)\, )\, n\cdot {J}(y) \ ,
\label{eq:Qgendef}
\EA 
where the vector $n_\mu$ selects the initial quantization
surface, which we take to be the null plane $\Sigma_n^\tau$. Therefore, the null-plane chiral symmetry charges are 
\begin{eqnarray}
{\tilde Q}_\alpha \ &=& \ \int\, d x^-\, d^2 \bit{x}_\perp\, {\tilde J}^+_{\alpha}(x^-, {\vec x}_\perp) \ ; \label{eq:npchargesgenA}\\
{\tilde Q}^5_\alpha(x^+) \ &=& \ \int\, d x^-\, d^2 \bit{x}_\perp\, {\tilde J}^+_{5\alpha}(x^-, {\vec x}_\perp, x^+) \ ,
\label{eq:npchargesgenB}
\end{eqnarray}
where the axial charges have been given explicit null-plane time dependence as they are not conserved
due to the explicit breaking operator in the Lagrangian.
These charges satisfy the $SU(N)_R\otimes SU(N)_L$ chiral algebra,
\BA
[\, {\tilde Q}^\alpha\,  ,\, {\tilde Q}^\beta\, ]\, =\, i\,f^{\alpha\beta\gamma} \, {\tilde Q}^\gamma
\ & , &\  
[\, {\tilde Q}_5^\alpha(x^+)\, ,\, {\tilde Q}^\beta\, ]\, =\,  i\,f^{\alpha\beta\gamma}\,  {\tilde Q}_5^\gamma(x^+) \ ;
\label{eq:LCalga}
\EA
\BA
[\, {\tilde Q}_5^\alpha(x^+) \, ,\, {\tilde Q}_5^\beta(x^+)\, ] \, =\,  i\,f^{\alpha\beta\gamma} \, {\tilde Q}^\gamma \ .
\label{eq:LCalgb}
\EA
We further assert that both types of chiral charges annihilate the vacuum. That is,
\BA
{\tilde Q}^\alpha\, |\, 0\,\rangle\, =\, {\tilde Q}_5^\alpha\, |\, 0\,\rangle\, =\, 0 \ .
\label{eq:vacuumgen}
\EA 
This is the statement that the front-form vacuum is invariant with
respect to the full $SU(N)_R\otimes SU(N)_L$ symmetry. In particular,
this implies that there can be no vacuum condensates that break
$SU(N)_R\otimes SU(N)_L$ on a null-plane. This may seem to be an odd
assumption, since the chiral charge is directly related to the
axial-vector current through eq.~\ref{eq:npchargesgenB}, and in
general one would expect that this current has a Goldstone boson pole
contribution, in turn implying that the chiral charges acting on the
vacuum state excite massless Goldstone bosons.  Below we will confirm the assertion,
eq.~\ref{eq:vacuumgen}, by using standard current-algebra polology to
show that indeed the Goldstone boson pole contribution to the null-plane
axial-vector current is absent.

\subsection{Symmetries of the reduced Hamiltonians}

\noindent Mixed commutators among the Poincar\'e generators and internal symmetry generators
can be expressed generally as~\cite{Feinberg:1973qb}:
\BA
\lbrack\, {Q}_\alpha(n\cdot x) \, ,\, P^\mu\rbrack \, &=&\, -i\;n^\mu  
\int\, d^4y\, \delta( n\cdot (x\, -\, y\,)\, )\, \partial_\nu J^\nu_\alpha(y)  \ ; \\
\lbrack\, {Q}_\alpha(n\cdot x) \, ,\, M^{\mu\nu}\rbrack \,&=&\,
i\int\, d^4y\, \delta( n\cdot (x\, -\, y\,)\, )\,\left(n^\mu y^\nu \ -\ n^\nu y^\mu \right) \partial_\kappa J^\kappa_\alpha(y)  \ .
\EA
From these expressions one then obtains the mixed commutator between the Pauli-Lubanski vector
and the internal symmetry charges:
\BA
\hskip-1.3em\lbrack{Q}_\alpha(n\cdot x)  , W_\nu\rbrack = \ft i2 \varepsilon_{\nu\delta\rho\sigma}
\int d^4y \delta( n\cdot (x - y))
\bigg\lbrack M^{\delta\rho}n^\sigma - \left(n^\delta y^\rho -n^\rho y^\delta \right)P^\sigma
\bigg\rbrack \partial_\kappa J^\kappa_\alpha(y).
\EA
Using these expressions, one finds the commutation relations between null-plane chiral charges and the
reduced Hamiltonians:
\BA
\lbrack\, {\tilde Q}^5_\alpha(x^+) \, ,\, M^2\rbrack \, &=&\, -2i\,P^+
\int\, d x^-\, d^2 \bit{x}_\perp\, \partial_\mu {\tilde J}^\mu_{5\alpha}(x^-, {\vec x}_\perp, x^+) \ ; \label{eq:mixedchiralc}\\
\lbrack\, {\tilde Q}^5_\alpha(x^+) \, ,\, M{\cal J}_r\rbrack \, &=&\, i\,\epsilon_{rs}\,P^+
\int\, d x^-\, d^2 \bit{x}_\perp\, {\Gamma}_s\,\partial_\mu {\tilde J}^\mu_{5\alpha}(x^-, {\vec x}_\perp, x^+) \ ,
\label{eq:mixedchirald}
\EA
where $\Gamma_s\equiv E_s - P^+ x_s $. Here and in what follows, we are assuming that $SU(N)_F$ is 
unbroken and therefore $\partial_\mu {\tilde J}^\mu_{\alpha}=0$ and
the reduced Hamiltonians commute with the $SU(N)_F$ charges:
\BA
\lbrack\, {\tilde Q}_\alpha\, ,\, M^2\rbrack &=& \lbrack\, {\tilde Q}_\alpha\, ,\, M{\cal J}_r\rbrack  \ = \ 0\ .
\label{eq:mixedvector}
\EA

\subsection{Goldstone's theorem on a null plane}

\noindent In the instant form, a symmetry has three possible fates in
the quantum theory: the symmetry remains exact and the current is
conserved, the symmetry is spontaneously broken and again the current
is conserved, or the symmetry is anomalous and the current is not
conserved. The front form realizes a fourth possibility: the symmetry
is spontaneously broken and the associated current in not conserved.
This fourth possibility is necessary in the front form because the
vacuum is invariant with respect to all internal symmetries.  In general, we
can write
\BA
\partial_\mu {\tilde J}^\mu_{5\alpha}(x^-, {\vec x}_\perp, x^+) \ = \ \epsilon_{\chi}\ {\tilde P}_{\alpha}(x^-, {\vec x}_\perp, x^+) \ ,
\label{eq:genpcac}
\EA
where $\epsilon_{\chi}$ is the parameter that gauges the amount of chiral symmetry breaking that is present in the Lagrangian.
Using the short hand,
\BA
|\,h\,\rangle\equiv \,|\, p^+\, ,\, {\vec p}_\perp\,;\, \lambda\,,\, h\,\rangle \ ,
\label{eq:HME2}
\EA for the momentum eigenstates, we take the matrix element of
eq.~\ref{eq:mixedchiralc} between momentum eigenstates, which gives
\BA
\langle\, h'\, |\lbrack\, {\tilde Q}^5_\alpha(x^+) \, ,\, M^2\rbrack |\, h\, \rangle\, &=&\, -2i\,p^+\,
\epsilon_{\chi}\, \int\, d x^-\, d^2 \bit{x}_\perp\, \langle\, h'\, | {\tilde P}_{\alpha}(x^-, {\vec x}_\perp, x^+) |\, h\, \rangle \ ; 
\label{eq:mixedchiral1}
\EA If the right hand side of this equation vanishes for all $h$ and
$h'$, then there can be no chiral symmetry breaking of any kind. 
Therefore, in order that the chiral symmetry be spontaneously broken, the
chiral current cannot be conserved and we have the following
constraint~\cite{Kim:1994rm,Yamawaki:1998cy} in the limit $\epsilon_\chi\rightarrow 0$:
\BA
\int\, d x^-\, d^2 \bit{x}_\perp\, \langle\, h'\, |{\tilde P}_{\alpha}(x^-, {\vec x}_\perp, x^+)|\, h\, \rangle &\longrightarrow & \frac{1}{\epsilon_\chi}\ +\ \ldots \ ,\label{eq:pizeromodea}
\EA
where the dots represent other possible terms that are non-singular in the limit $\epsilon_\chi\rightarrow 0$.
Now we will show that this condition implies the existence of $N^2-1$ Goldstone bosons~\footnote{Note that if we took eq.~\ref{eq:pizeromodea}
as a constraint on the operator ${\tilde P}_\alpha$ rather than on its matrix elements, then this constraint would be viewed
as a constraint on the zero-modes of the operator~\cite{Kim:1994rm,Yamawaki:1998cy}. Here we work entirely with matrix elements.}. 
We will assume that ${\tilde P}_\alpha$ is an interpolating operator for Lorentz-scalar fields $\phi_\alpha^i$, and 
therefore we can write
\BA
{\tilde P}_\alpha(x) \ =\ \sum_i {\cal Z}_i\,\phi_\alpha^i(x) \ 
\label{eq:zfactors}
\EA
where the ${\cal Z}_i$ are overlap factors. Using the reduction
formula we relate the matrix elements of field operators between
physical states to transition amplitudes. Of course here it is understood
that there is no selection rule which would forbid these transitions.
The S-matrix element for the transition $h(p)\rightarrow h'(p')+\phi^i_\alpha(q)$ can be defined by
\BA
\langle \, h'\,;\, \phi_\alpha^i(q)\, |
S |\, h\, \rangle  &\equiv&  i(2\pi)^4\,\delta^4(\, p\,-\, p'-q)\,
{\cal M}^i_\alpha (\,p',\,\lambda',\, h'\,;\,p,\,\lambda,\, h\,) \nn \\
&=& i\int d^4x\,e^{-iq\cdot x}\,\left(-q^2+M_{\phi^i}^2\right)\,\langle\, h'\, |\, \phi_\alpha^i(x)\,|\, h\, \rangle
\label{eq:HME3}
\EA
where ${\cal M}^i_\alpha$ is the Feynman amplitude and in the second line we have used
the reduction formula. It then follows that
\BA
\langle\, h'\, |\, \phi_\alpha^i(x)\,|\, h\, \rangle & =& 
-e^{iq\cdot x}\,{1\over{q^2-M_{\phi^i}^2}}\, {\cal M}^i_\alpha(q) \ .
\label{eq:reductiongen}
\EA
Using this formula together with eq.~\ref{eq:zfactors} in eq.~\ref{eq:mixedchiral1} then gives
\BA
\langle\, h'\, |\lbrack\, {\tilde Q}^5_\alpha(x^+) \, ,\, M^2\rbrack |\, h\, \rangle &=& 2i\,p^+\,
(2\pi)^3\,\delta(\,q^+\,)\,\delta^2(\,{\vec q}_\perp \,)\,e^{ix^+q^-}\sum_i \frac{\epsilon_\chi\, {\cal Z}_i}
{2q^+q^-\,-\,{\vec q}_\perp^{\;2}-M_{\phi^i}^2}\,{\cal M}^i_\alpha(q) \nn \\ 
&=&{} -2i\,p^+\, (2\pi)^3\,\delta(\,q^+\,)\,\delta^2(\,{\vec q}_\perp \,)\,e^{ix^+q^-}\sum_i \frac{\epsilon_\chi\, {\cal Z}_i}
{M_{\phi^i}^2}\,{\cal M}^i_\alpha(q^-) \ ,
\label{eq:mixedchiralproof}
\EA 
where in the second line we have used the momentum delta
functions.  In order that the right hand side not vanish in the
symmetry limit, there must be at least one field $\phi_\alpha^i$ whose
mass-squared vanishes proportionally to $\epsilon_\chi$ as
$\epsilon_\chi\rightarrow 0$. We will denote this field as
$\pi^\alpha\equiv \phi_\alpha^1$ with 
\BA M_\pi^2 \ =\ c_p\,
\epsilon_\chi \ ,
\label{eq:pionmass}
\EA 
where $c_p$ is a constant of proportionality. There are therefore
$N^2-1$ massless fields $\pi_\alpha$ in the symmetry limit, which we
identify as the Goldstone bosons.  It is noteworthy that this proof
relies entirely on physical matrix elements; i.e. there is no need to
assume the existence of a vacuum condensate that breaks the chiral
symmetry. Of course, in instant-form QCD, we know that the
proportionality constant in eq.~\ref{eq:pionmass} contains the quark
condensate. This issue will be resolved below in
section~\ref{sec:qcd}.  While we have carried out this proof in the
case of $SU(N)_R\otimes SU(N)_L$ broken to $SU(N)_V$, it is clearly
easily generalized to other systems.

We can now write ${\tilde P}_\alpha \ =\ {\cal Z}\,\pi_\alpha\ +\ \ldots$
where the dots represent non-Goldstone boson fields, and 
\BA
\langle\, h'\, |\,  \partial_\mu\,{\tilde J}^\mu_{5\alpha}(x)\,|\, h\, \rangle & =& 
\langle\, h'\, |\,  {\bar{\cal Z}}\,M_\pi^2\,\pi_\alpha(x)\,|\, h\, \rangle \ ,
\label{eq:pcacgen}
\EA
where ${\bar{\cal Z}}\equiv{\cal Z}/c_p$. Here, as in the usual current algebra
manipulations, we have assumed that only the Goldstone bosons couple to the
axial-vector current, and it is now a standard exercise to determine the overlap factor. First, define
the Goldstone-boson decay constant, $F_\pi$, via
\BA
\langle\, 0\, |\,{\tilde J}^\mu_{5\alpha}(x)\,|\, \pi_\beta\, \rangle & \equiv& -i\,p^\mu\;F_\pi\,\delta_{\alpha\beta}\;e^{ip\cdot x} \ ,
\label{eq:piondecaydef}
\EA
where $|\, \pi_\beta\, \rangle \equiv |\, p^+\, ,\, {\vec p}_\perp\,;\, 0,,\, \pi_\beta\, \rangle$.
Taking the divergence of the current and raising eq.~\ref{eq:pcacgen} to an operator relation yields
\BA
\langle\, 0\, |\,{\bar{\cal Z}}\,M_\pi^2\,\pi_\alpha(x) \,|\, \pi_\beta\, \rangle & =& F_\pi\,M_\pi^2\,\delta_{\alpha\beta}\;e^{ip\cdot x} \ .
\label{eq:piondecaydef2}
\EA
The normalization of the Goldstone-boson field,
\BA
\langle\, 0\, |\,\pi_\alpha(x)\,|\, \pi_\beta\, \rangle &= & \delta_{\alpha\beta}\;e^{ip\cdot x} \ ,
\label{eq:pionfieldnorm}
\EA
then gives $\bar{\cal Z}=F_\pi$ and we recover the standard operator relation
\BA
\partial_\mu\,{\tilde J}^\mu_{5\alpha}(x) & =&  F_\pi\,M_\pi^2\,\pi_\alpha(x) \ .
\label{eq:pcacgen2}
\EA
We can now express the mixed Lie bracket, eq.~\ref{eq:mixedchiralproof}, as 
\BA
\langle\, h'\, |\lbrack\, {\tilde Q}^5_\alpha(x^+) \, ,\, M^2\rbrack |\, h\, \rangle 
&=& -2i\,p^+\, (2\pi)^3\,\delta(\,q^+\,)\,\delta^2(\,{\vec q}_\perp \,)\,e^{ix^+q^-}F_\pi\,{\cal M}_\alpha(q^-) \ ,
\label{eq:mixedfinal}
\EA
where here ${\cal M}_\alpha(q^-)$ is the Feynman amplitude for the transition $h(p)\rightarrow h'(p')+\pi_\alpha(q)$.
We see that while the chiral current is not conserved, its divergence is proportional to an S-matrix element.
Noting that 
\BA
\hskip-1.3em\langle\, h'\, |\lbrack\, {\tilde Q}^5_\alpha(x^+) \, ,\, M^2\rbrack |\, h\, \rangle 
&=& 2p^+q^-\langle\, h'\, |{\tilde Q}^5_\alpha(x^+) |\, h\, \rangle \ =\  -2p^+\langle\, h'\, |i\,\frac{d}{dx^+}{\tilde Q}^5_\alpha(x^+) |\, h\, \rangle \ ,
\label{eq:mixeddef}
\EA
and from the definition of the chiral charge, eq.~\ref{eq:npchargesgenB},
\BA
\langle\, h'\, |{\tilde Q}^5_\alpha(x^+) |\, h\, \rangle &=& 
(2\pi)^3\,\delta(\,q^+\,)\,\delta^2(\,{\vec q}_\perp \,)\,e^{ix^+q^-}
\langle\, h'\, |{\tilde J}^+_{5\alpha} (0) |\, h\, \rangle \ ,
\label{eq:chiralchargecurrent}
\EA
one finds, using eq.~\ref{eq:mixedfinal},
\BA
{\cal M}_\alpha(q^-)
& =& 
\frac{i\,q^-}{F_\pi}\, \langle\, h'\, |\,  {\tilde J}^+_{5\alpha}(0)\,|\, h\, \rangle \ ,
\label{eq:matcurr}
\EA
or, in Lorentz-invariant form,
\BA
{\cal M}_\alpha(q) & =& 
\frac{i\ q_\mu}{F_\pi}\, \langle\, h'\, |\,  {\tilde J}^\mu_{5\alpha}(0)\,|\, h\, \rangle \ ,
\label{eq:HME8a}
\EA
which is the standard current-algebra result.  In order to confirm
some of these properties in a better-known fashion, and to address the
assumption we have made that the chiral charges annihilate the vacuum,
we will now consider current algebra polology on the
null-plane.

%
\subsection{Polology and the chiral invariant vacuum}

\begin{figure}[!t]
  \centering
     \includegraphics[scale=0.25]{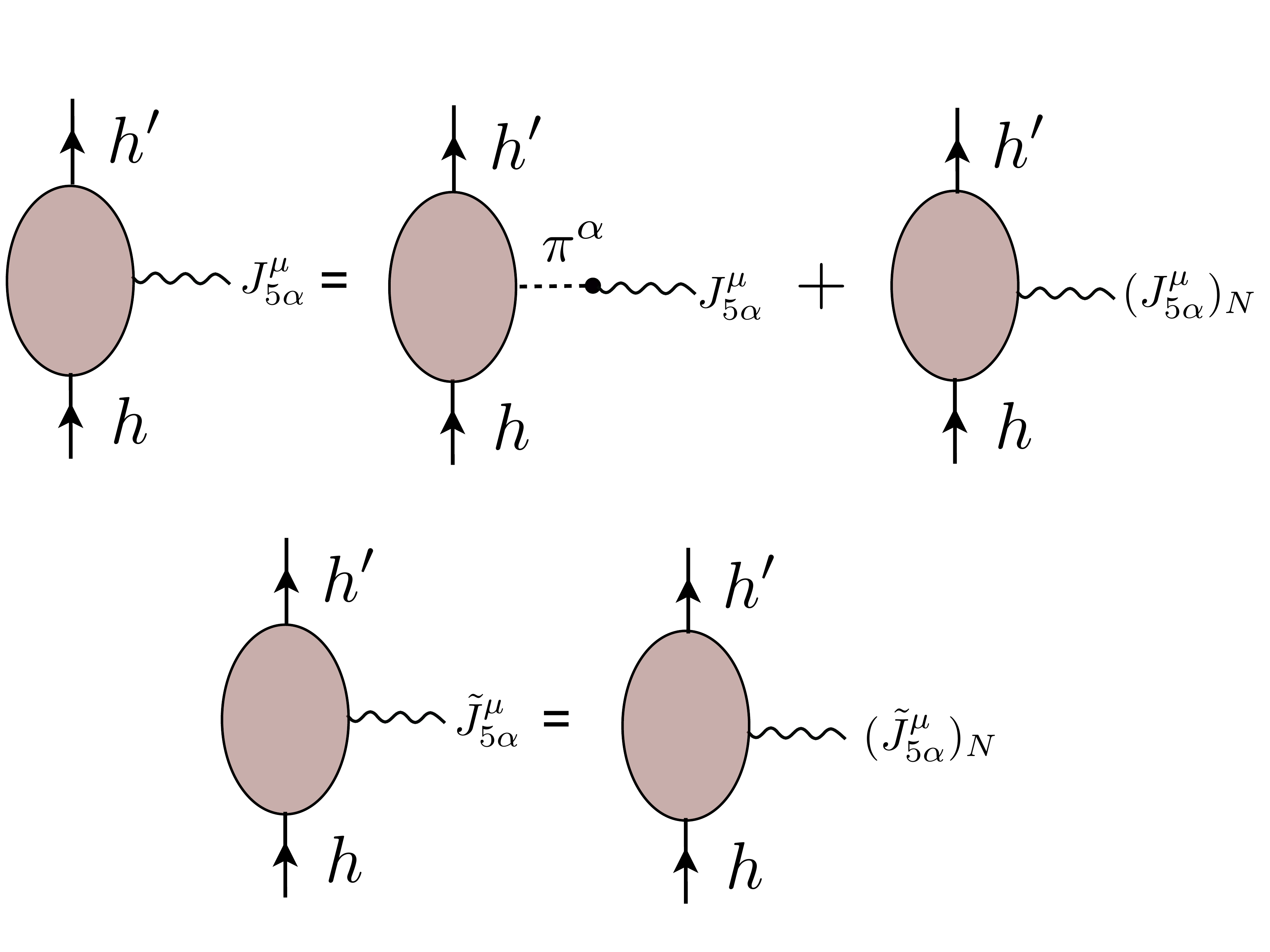}
     \caption{Above shows the standard instant-form polology; the matrix element of the chiral current
has a Goldstone-boson pole piece, and a non-pole piece. These two contributions cancel in the symmetry limit
ensuring a conserved chiral current. Below shows the standard front-form polology; the Goldstone-boson pole contribution
is absent and therefore the current is not conserved but rather has a divergence which is proportional to the matrix element
for the emission or absorption of a Goldstone boson.}
  \label{fig:polo}
\end{figure}

\noindent  Our starting point is the matrix element between hadronic states $h$
and $h'$ of the axial-vector current, which can be written in a general
way as~\cite{Weinberg:1995mt,Weinberg:1996kr}
\BA
\langle\, h'\, |\,  {\tilde J}^\mu_{5\alpha}(0)\,|\, h\, \rangle \, =\, 
\frac{i F_\pi\, q^\mu}{q^2-M_\pi^2}\, {\cal M}_\alpha
\, +\, 
\langle\, h'\, |\,  {\tilde J}^\mu_{5\alpha}(0)\,|\, h\, \rangle_{N}
\label{eq:HME1}
\EA
where as before $q=p-p^\prime$. Using translational invariance, we have
\BA
\langle\, h'\, |\,  {\tilde J}^\mu_{5\alpha}(x)\,|\, h\, \rangle \, =\, 
e^{iq\cdot x}\,\langle\, h'\, |\,  {\tilde J}^\mu_{5\alpha}(0)\,|\, h\, \rangle \ .
\label{eq:HME4}
\EA
It follows that 
\BA
\langle\, h'\, |\,  \partial_\mu\,{\tilde J}^\mu_{5\alpha}(x)\,|\, h\, \rangle & =& 
i\, q_\mu\,\langle\, h'\, |\,  {\tilde J}^\mu_{5\alpha}(x)\,|\, h\, \rangle  \nonumber \\
& =& 
e^{iq\cdot x}\, \Bigl\lbrack\,
\frac{-F_\pi\, q^2}{q^2-M_\pi^2}\, {\cal M}_\alpha
\, +\, 
i\ q_\mu\, \langle\, h'\, |\,  {\tilde J}^\mu_{5\alpha}(0)\,|\, h\, \rangle_{N}\, \Bigr\rbrack \ ,
\label{eq:HME5}
\EA
and using
\BA
\langle\, h'\, |\,  \partial_\mu\,{\tilde J}^\mu_{5\alpha}(x)\,|\, h\, \rangle & =& 
\langle\, h'\, |\,  F_\pi\,M_\pi^2\,\pi_\alpha(x)\,|\, h\, \rangle \ ,
\label{eq:HME6}
\EA
and the reduction formula, eq.~\ref{eq:reductiongen}, reproduces eq.~\ref{eq:HME8a}.
Note that in null-plane coordinates eq.~\ref{eq:HME1} gives
\BA
\langle\, h'\, |\,  {\tilde J}^+_{5\alpha}(0)\,|\, h\, \rangle \, =\, 
\frac{i F_\pi\, q^+}{2q^+q^-\,-\,{\vec q}_\perp^{\;2}-M_\pi^2}\, {\cal M}_\alpha
\, +\, 
\langle\, h'\, |\,  {\tilde J}^+_{5\alpha}(0)\,|\, h\, \rangle_{N} \ .
\label{eq:LCC2}
\EA
We therefore have
\BA
\lim_{{q+,{\vec q}_\perp}\rightarrow 0}\langle\, h'\, |\,  {\tilde J}^+_{5\alpha}(0)\,|\, h\, \rangle \ =\ 
\langle\, h'\, |\,  {\tilde J}^+_{5\alpha}(0)\,|\, h\, \rangle_{N} \ .
\label{eq:LCC3}
\EA
By comparing with eq.~\ref{eq:chiralchargecurrent}, it is clear that
the null-plane chiral charges, by construction, do not excite the
Goldstone boson states. The property, eq.~\ref{eq:vacuumgen}, of
vacuum annihilation which we assumed above, is therefore a general
property of the null-plane chiral charges.

Again consider the space-integrated current divergence in the front-form, but now using eq.~\ref{eq:HME5}. One finds
\BA
&&\int\, d x^-\, d^2 {\bf x_\perp}\,\langle\, h'\, |\,  \partial_\mu\,{\tilde J}^\mu_{5\alpha}(x)\,|\, h\, \rangle \  = \
(2\pi)^3\,\delta(\,q^+\,)\,\delta^2(\,{\vec q}_\perp \,)\,e^{ix^+q^-}
\,\langle\, h'\, |\,  \partial_\mu\,{\tilde J}^\mu_{5\alpha}(0)\,|\, h\, \rangle  \nonumber \\
&&\qquad = 
(2\pi)^3\,\delta(\,q^+\,)\,\delta^2(\,{\vec q}_\perp \,)\,e^{ix^+q^-}\,
\Bigl\lbrack\,
\frac{-F_\pi\, \left( 2q^+q^-\,-\,{\vec q}_\perp^{\;2} \right)}{2q^+q^-\,-\,{\vec q}_\perp^{\;2}-M_\pi^2}\, {\cal M}_\alpha
\, +\, 
i\ q_\mu\, \langle\, h'\, |\,  {\tilde J}^\mu_{5\alpha}(0)\,|\, h\, \rangle_{N} \,
\Bigl\rbrack \nonumber \\
&&\qquad =
(2\pi)^3\,\delta(\,q^+\,)\,\delta^2(\,{\vec q}_\perp \,)\,e^{ix^+q^-}
\, i\ q_\mu\, \langle\, h'\, |\,  {\tilde J}^\mu_{5\alpha}(0)\,|\, h\, \rangle_{N} \nonumber \\
&&\qquad =
(2\pi)^3\,\delta(\,q^+\,)\,\delta^2(\,{\vec q}_\perp \,)\,e^{ix^+q^-}
\, F_\pi\,{\cal M}_\alpha(q^-) \ ,
\label{eq:HME9a}
\EA
where in the third line the momentum delta functions have been used,
and in the last line we have used eq.~\ref{eq:HME8a} and eq.~\ref{eq:LCC3}. Now using
eq.~\ref{eq:mixedchiralc}, we see that we have recovered
eq.~\ref{eq:mixedfinal}. In this derivation we see explicitly that the
Goldstone-boson pole does not contribute to the divergence of the
axial-current.  It is for this reason that the current cannot be
conserved.  For purposes of comparison, recall that in the instant
form, one has
\BA
\int\, d^3 x\,\langle\, h'\, |\,  \partial_\mu\,{J}^\mu_{5\alpha}(x)\,|\, h\, \rangle & =& 
(2\pi)^3\,\delta^3(\,{\vec q}\,)
\,\langle\, h'\, |\,  \partial_\mu\,{J}^\mu_{5\alpha}(0)\,|\, h\, \rangle \ ; \nonumber \\
& =& 
(2\pi)^3\,\delta^3(\,{\vec q}\,)\,
\Bigl\lbrack\,
\frac{-F_\pi\, q_0^2}{q_0^2-M_\pi^2}\, {\cal M}_\alpha
\, +\, 
i\ q_\mu\, \langle\, h'\, |\,  {J}^\mu_{5\alpha}(0)\,|\, h\, \rangle_{N} \,
\Bigl\rbrack \ ; \nonumber \\
& \mapright{M_\pi\rightarrow 0}& 
(2\pi)^3\,\delta^3(\,{\vec q}\,)\,
\Bigl\lbrack\,
-F_\pi \, {\cal M}_\alpha
\, +\, 
i\ q_\mu\, \langle\, h'\, |\,  {J}^\mu_{5\alpha}(0)\,|\, h\, \rangle_{N} \,
\Bigr\rbrack \ ; \nonumber \\
& =& 0 \ ,
\label{eq:HME9}
\EA
where in the last line, eq.~\ref{eq:HME8a} has once again been used.
Here there is a cancellation between the pole and non-pole parts of the matrix element
which ensure that the integrated current divergence vanishes in the chiral limit.
This analysis, which is expressed pictorially in fig.\ref{fig:polo},  suggests that the front-form and instant-form  axial-vector currents
are related, at the operator level, through
\BA
{\tilde J}^\mu_{5\alpha}\ =\ {J}^\mu_{5\alpha}\ -\ ({J}^\mu_{5\alpha})_{\it GB\,pole}
\label{eq:HME11}
\EA 
where the second term on the right is the purely Goldstone-boson pole part of the axial-vector current.
We will see that this peculiar realization of chiral symmetry does indeed emerge in QCD.

It is useful to define new objects which give a matrix-element representation of the 
internal-symmetry charges~\cite{Weinberg:1969hw,Weinberg:1969db}:
\BA
\langle\, h'\, |\, {\tilde Q}^5_\alpha(x^+) \,|\, h\, \rangle \, =\, 
(2\pi)^3\,2\,p^+\,\delta(\,q^+\,)\,\delta^2(\,{\vec q}_\perp \,)
\lbrack\, X_\alpha(\lambda) \,\rbrack_{h' h}\,
\delta_{\lambda'\lambda} \ ;
\label{eq:HME40}
\EA
\BA
\langle\, h'\, |\, {\tilde Q}_\alpha \,|\, h\, \rangle \, =\, 
(2\pi)^3\,2\,p^+\,\delta(\,q^+\,)\,\delta^2(\,{\vec q}_\perp \,)
\lbrack\, T_\alpha \,\rbrack_{h}\, \delta_{h h'}\,
\delta_{\lambda'\lambda} \ . 
\label{eq:HME41}
\EA
These definitions are particularly useful as they allow the preservation of the Lie-algebraic structure of the operator algebra in the
case where correlation functions are given purely by single-particle states. The matrix element for Goldstone boson emission and absorption is:
\BA
{\cal M}_\alpha (\,p',\,\lambda',\, h'\,;\,p,\,\lambda,\, h\,) \ =\ \frac{i}{F_\pi}\,
(\,M^2_h\ -\ M^2_{h'}\,)\, \lbrack\, X_\alpha(\lambda) \,\rbrack_{h' h}\,
\delta_{\lambda'\lambda} \ . 
\label{eq:HME42}
\EA
As one might expect, in the limit that chiral symmetry is restored through a second-order phase transition,
the matrix $\lbrack\, X_\alpha(\lambda) \,\rbrack_{h' h}$ becomes a true symmetry generator~\cite{Weinberg:1995fe}. 
In this limit, one also expects that the states $h'$ and $h$ become degenerate. In order that the matrix element of
eq.~\ref{eq:HME42} not vanish in this limit, ${F_\pi}$ must approach zero in the symmetry limit in precisely the same way~\cite{Weinberg:1995fe}. 
The role of ${F_\pi}$ as an order parameter of chiral symmetry breaking is then apparent in eq.~\ref{eq:mixedfinal},
as the mixed-Lie bracket vanishes as ${F_\pi}\rightarrow 0$. Therefore, ${F_\pi}$ is an order parameter of chiral 
symmetry breaking on the null-plane.

%
\subsection{Broken chiral symmetry and spin}
\label{susec:spin}

\noindent Using the results of the previous two sections one finds
\BA
&&\hspace{-.4in}\langle\, h'\,,\,\lambda' |\lbrack\, {\tilde Q}^5_\alpha(x^+) \, ,\, M^2\rbrack |\, h\,,\,\lambda \rangle = 
\delta_{\lambda',\lambda}(2\pi)^3\,2\,p^+\,\delta(q^+)\delta^2({\vec q}_\perp)
\left(M_h^2-M_{h'}^2\right)\lbrack\, X_\alpha(\lambda) \,\rbrack_{h' h}
\label{eq:mixedchiral1out}
\EA
and
\BA
&&\hspace{-.92in}\langle\, h'\,,\,\lambda' |\lbrack\, {\tilde Q}^5_\alpha(x^+) \, ,\, M{\cal J}_\pm\rbrack |\, h\,,\,\lambda \rangle\ =\ 
\delta_{\lambda',\lambda\pm 1}\;(2\pi)^3\,2\,p^+\,\delta(\,q^+\,)\,\delta^2(\,{\vec q}_\perp \,)\nonumber\\
&&\times\;\bigg\lbrack M_h\,c_\pm(h,\lambda)\,\lbrack\, X_\alpha(\lambda\pm 1) \,\rbrack_{h' h}\,-\,
M_h'\,c_\mp(h',\lambda')\,\lbrack\, X_\alpha(\lambda) \,\rbrack_{h' h}\bigg\rbrack,
\label{eq:mixedchiral2out}
\EA
where ${\cal J}_\pm\equiv{\cal J}_1\pm i{\cal J}_2$ and $c_\pm(h,\lambda)\equiv\sqrt{j_h(j_h+1)-\lambda(\lambda\pm1)}$.
Eq.~\ref{eq:mixedchiral2out} has been obtained by a direct evaluation of the left-hand side using the usual
angular momentum ladder relations and eq.~\ref{eq:HME40}.
Written in this form, it is clear that in the presence of spontaneous symmetry breaking, the mixed Lie brackets between
the reduced Hamiltonians and the chiral charge are directly related to Goldstone-boson transition amplitudes and are
non-vanishing in the symmetry limit. The spin reduced Hamiltonians imply constraints on Goldstone-boson transitions that 
change the helicity by one unit.

An important consequence of eqs.~\ref{eq:mixedchiral1out} and ~\ref{eq:mixedchiral2out} which will 
prove useful below is that chiral symmetry breaking remains relevant even when there are no mass 
splittings. If we take $M_h\;=\;M_{h'}$, then chiral symmetry breaking arises solely through the
transverse spin operator, ${\cal J}_r$, which is dynamical on the null-plane. That is,
\BA
&&\hspace{-.92in}\langle\, h'\,,\,\lambda' |\lbrack\, {\tilde Q}^5_\alpha(x^+) \, ,\, {\cal J}_\pm\rbrack |\, h\,,\,\lambda \rangle\ =\ 
\delta_{\lambda',\lambda\pm 1}\;(2\pi)^3\,2\,p^+\,\delta(\,q^+\,)\,\delta^2(\,{\vec q}_\perp \,)\nonumber\\
&&\times\;\bigg\lbrack \,c_\pm(h,\lambda)\,\lbrack\, X_\alpha(\lambda\pm 1) \,\rbrack_{h' h}\,-\,
\,c_\mp(h',\lambda')\,\lbrack\, X_\alpha(\lambda) \,\rbrack_{h' h}\bigg\rbrack \ .
\label{eq:mixedchiral2outb}
\EA
In this case, Goldstone's theorem must be obtained from the relation
\BA
\hspace{-.32in}\langle\, h'\, |\;M\;\lbrack\, {\tilde Q}^5_\alpha(x^+) \, ,\, {\cal J}_r\rbrack |\, h\, \rangle\, &=&\, -i\,\epsilon_{rs}\,p^+\,
\epsilon_{\chi}\, \int\, d x^-\, d^2 \bit{x}_\perp\, \langle\, h'\, |{x}_s\,  {\tilde P}_{\alpha}(x^-, {\vec x}_\perp, x^+) |\, h\, \rangle  ,
\label{eq:mixedchiral2}
\EA
and its corresponding constraint
\BA
\int\, d x^-\, d^2 \bit{x}_\perp\, \langle\, h'\, |{x}_s\,{\tilde P}_{\alpha}(x^-, {\vec x}_\perp, x^+)|\, h\, \rangle &\longrightarrow & \frac{1}{\epsilon_\chi}\ +\ \ldots 
\label{eq:pizeromodeb}
\EA
in the symmetry limit, ${\epsilon_\chi}\rightarrow 0$.
Following the same steps as for the mass-squared reduced Hamiltonian, we have
\BA
\hskip-1.8em\langle\, h'\, |\lbrack\, {\tilde Q}^5_\alpha(x^+) \, ,\, {\cal J}_r\rbrack |\, h\, \rangle
= 
-\epsilon_{rs}\,\frac{p^+}{M_h}\, (2\pi)^3\,\delta(\,q^+\,)\,\delta^2(\,{\vec q}_\perp \,)\,e^{ix^+q^-}\sum_i \frac{\epsilon_\chi\, {\cal Z}_i}
{M_{\phi^i}^2}\left(\frac{\partial}{\partial q_s}\;{\cal M}^i_\alpha(q)\right)
\label{eq:mixedchiralproof2}
\EA which again leads, via the same logic presented above, to
Goldstone's theorem. Therefore, even if $M^2$ commutes with the chiral
charges, the chiral symmetry breaking contained in the spin
Hamiltonians implies the presence of massless states. Evaluating
eq.~\ref{eq:mixedchiralproof2} in the rest frame, where
$p^+\rightarrow M_h/\sqrt{2}$ and ${\cal J}_r\rightarrow{J_r}$, and using eq.~\ref{eq:HME8a}, gives
\BA
\langle\, h'\, |\lbrack\, {\tilde J}^+_{5\alpha}\, ,\, {J}^r\rbrack |\, h\, \rangle
&=& 
i\frac{1}{\sqrt{2}}\epsilon^{rs}\langle\, h'\, |\; {\tilde J}^s_{5\alpha}\; |\, h\, \rangle \ ,
\label{eq:wetheorem}
\EA
which is simply the statement that the axial current transform as a vector operator.

\subsection{General operator algebra and the chiral basis}

\noindent A physical system with an $SU(N)_R\otimes SU(N)_L$ chiral
symmetry broken to the vector subgroup $SU(N)_F$ may be expressed as a
dynamical Hamiltonian system which evolves with null-plane time,
whose reduced Hamiltonians satisfy the $U(2)$ algebra of eq.~\ref{eq:dynalgarbB},
and in addition have non-vanishing Lie brackets with the non-conserved
chiral charges. In operator form the reduced Hamiltonians satisfy:
\BA
[\, {\tilde Q}^\beta_5(x^+)\, ,\, M^2\, ] \ \neq \ 0 \quad ; \quad
[\, {\tilde Q}^\beta_5(x^+)\, ,\, M {\cal J}_\pm\, ]\ \neq \ 0 \ , 
\label{eq:Qcomm2agen}
\EA
which express the spontaneous breaking of the chiral symmetry. 
Eq.~\ref{eq:Qcomm2agen} has the general operator solution
\BA
M^2\, &=&\, M_{\bf 1}^2\, +\, \sum_{\cal R}\, M_{\cal R}^2 \ \ ; \nonumber \\
M {\cal J}_\pm\, &=&(M {\cal J}_\pm)\, _{\bf 1}\, +\, \sum_{\cal R}\, (M {\cal J}_\pm)\, _{\cal R}
\label{eq:MESONScommgen}
\EA 
where ${\bf 1}$ denotes the singlet $SU(N)_R\otimes SU(N)_L$
representation, $({\bf 1},{\bf 1})$, and ${\cal R}=({\bf {\cal
    R}_{R}},{\bf {\cal R}_{L}})$ is a non-trivial representation. Note
that all three symmetry-breaking reduced Hamiltonians must transform
in the same way. This follows directly from eqs.~\ref{eq:mixedchiralc}
and~\ref{eq:mixedchirald}.  

It is useful to give a heuristic description of the consequences of
this algebraic structure.  Consider an interpolating field operator,
$a_h^\dagger$ which creates a momentum state $h$ out of the vacuum;
that is,
\BA
a_h^\dagger\, |\, 0\,\rangle\, =\, |\, h\,\rangle\, \ .
\label{eq:hdef}
\EA
Here and below for simplicity we will suppress the flavor indices.
Because the null-plane chiral charges annihilate the vacuum, ${\tilde Q}_5\, |\, 0\,\rangle\, =\, 0 $,
one has
\BA
{\tilde Q}_5 \, |\, h\,\rangle\, &=&\, 
\lbrack\, {\tilde Q}_5 \, ,\, a_h^\dagger\, \rbrack \, |\, 0\,\rangle\ . 
\label{eq:htransf}
\EA
Now we will assume that the interpolating field operator $a_h^\dagger$ has definite chiral transformation properties with respect 
to the chiral charge in the sense that 
\BA
\lbrack\, {\tilde Q}_5 \, ,\, a_h^\dagger\, \rbrack \, =\, C'\, a_{h'}^\dagger\, +\, C''\, a_{h''}^\dagger\, +\, \ldots \ ,
\label{eq:htransfprops}
\EA
where $C',C'',\ldots$ are group-theoretic factors.
This is simply the statement that the field operators $\lbrace a_h, a_{h'},a_{h''},\ldots\rbrace$ are in a non-trivial $SU(N)_R\otimes SU(N)_L$ representation. It
then follows from eq.~\ref{eq:htransf} that
\BA
{\tilde Q}_5 \, |\, h\,\rangle\, &=&\, C'\,|\, h'\,\rangle\, + C''\,|\, h''\,\rangle\, +\,\ldots \ ,
\label{eq:htransf2}
\EA 
and therefore the states $\lbrace h, {h'},{h''},\ldots\rbrace$ are also in an $SU(N)_R\otimes SU(N)_L$ representation~\footnote{
Note that the instant-form interpolating operators also fill out 
$SU(N)_R\otimes SU(N)_L$ representations. However, the instant-form charges do not annihilate the vacuum,
  i.e. $Q_5|\,0\, \rangle \equiv |\, {\bf\omega}\, \rangle$, it follows
  that ${Q}_5\, |\, h\,\rangle\, =\, \lbrack\, {Q}_5\, ,\,
  a_h^\dagger\, \rbrack \, |\, 0\,\rangle\, +\, |\, h\,;\, {\bf\omega}
  \,\rangle$.  Therefore ${Q}^5 \, |\, h\,\rangle\, =\, C'\,|\,
  h'\,\rangle\, + C''\,|\, h''\,\rangle\, +\,\ldots\,+ \, |\, h\,;\,
  {\bf\omega} \,\rangle$ and the utility of chiral symmetry as a
  classification symmetry is lost.}.

One then has, for instance, 
\BA
\langle\,h'\,|\,{\tilde Q}_5\, |\, h\,\rangle & =& C'  \ ; \\
\langle\,h''\,|\,{\tilde Q}_5\, |\, h\,\rangle &=& C'' \ ,
\label{eq:piontrans}
\EA 
which are, via eq.~\ref{eq:HME40}, Goldstone-boson transition matrix elements.
If $\lbrace h, h',h'',\ldots\rbrace$ are in an irreducible
representation, then the $C$'s are completely determined by the
symmetry (i.e. are Clebsch-Gordon coefficients), while if the
representation is reducible, then the $C$'s will depend on the mixing
angles which mix the irreducible representations.  Therefore through
the study of Goldstone-boson transitions one learns about the chiral
representations filled out by the physical states~\footnote{Here it should be
stressed that the chiral multiplet structure of the states is useful
only when the null-plane chiral charges mediate transitions
between single-particle states~\cite{Weinberg:1969hw,Weinberg:1969db}. Multi-particle
states obscure the algebraic consequences of null-plane chiral symmetry.}.
To learn more about the chiral representations, one considers the
mixed Lie brackets, eqs.~\ref{eq:mixedchiral1out} and \ref{eq:mixedchiral2out}.
Knowledge of the transformation properties of
the chiral-symmetry breaking reduced Hamiltonians gives
information about how the hadron masses and spins are related, and therefore in
how the irreducible representations mix with each other when the
symmetry is broken. 

A natural null-plane basis can be written as
\BA
|\, k^+\, ,\, {\vec k}_\perp\, ;\, \lambda\,, h\, ,\, ({\bf {\cal R}_{R}},{\bf {\cal R}_{L}})\, \rangle \ .
\label{eq:HBpluschiraldefined}
\EA 
While the mass eigenstates are eigenstates of helicity, they clearly
are not eigenstates of $SU(N)_R\otimes SU(N)_L$ when the symmetry is
spontaneously broken. Nevertheless, the chiral basis is useful when
the state $h$ can only appear in a finite number of chiral
representations, even though $h$ may be in an infinite-dimensional
reducible chiral representation, as is the case generally in QCD at large-$\nc$~\cite{'tHooft:1973jz,Weinberg:1990xn}. In the
chiral basis, the reduced Hamiltonian matrix $M^2$ is then of finite
rank, even though there can be submatrices of infinite rank (and
therefore the Fock expansion in the number of constituents is
infinite). Ultimately, the utility of the chiral basis is
determined by comparison with experiment~\cite{Adler:1965ka,Weisberger:1965hp,Gilman:1967qs,Weinberg:1969hw,Weinberg:1969db,Weinberg:1990xn,Weinberg:1994tu,Beane:1999hp,Beane:2002ud,Beane:2002td}.

%
\section{QCD in the front form}
\label{sec:qcd}

\subsection{Basic instant-form conventions}

\noindent In this section, we will review the relevant symmetry
properties of the instant-form QCD Lagrangian for purposes of
establishing conventions which will clarify the null-plane
description. Consider the QCD Lagrangian with $N$ flavors of light
quarks and $N_c$ colors:
\begin{equation}
{\cal L}_{\scriptstyle\rm QCD} (x)
=
 \bar\psi (x)\Big\lbrack\ft{i}2\left(
\stackrel{{}_\rightarrow}{D}_\mu - \stackrel{{}_\leftarrow}{D}_\mu
\right)
\gamma^\mu  - \mathbb{M} \Big\rbrack\psi (x)
-
\ft14 F_{\mu\nu}^a (x) F^{\mu\nu}_a (x)
\,
\label{QCDET}
\end{equation}
where $\mathbb{M}$ is the quark mass matrix, for now taken as a diagonal $N\times N$ matrix, and
the covariant derivatives are 
\begin{equation}
\stackrel{{}_\rightarrow}{D}_\mu
\, = \,
\stackrel{{}_\rightarrow}{\partial}_\mu
- \, i g \, t^a A_\mu^a (x)
\, , \qquad
\stackrel{{}_\leftarrow}{D}_\mu
\, = \,
\stackrel{{}_\leftarrow}{\partial}_\mu
+ \, i g \, t^a A_\mu^a (x)
\, ,
\end{equation}
where $g$ is the strong coupling constant, and indices $a,b,\ldots$ are taken as adjoint indices of the $SU(3)$-color gauge group.
The Lagrangian is invariant with respect to baryon number and 
singlet axial transformations
\begin{equation}
\psi\rightarrow e^{-i\theta} \psi \ \ \ ,\ \ \  \psi\rightarrow e^{-i\theta \gamma_5} \psi \ ,
\end{equation}
with associated currents
\begin{equation}
{J}^\mu\ =\ \bar\psi \gamma^\mu\psi \ \ \ , \ \ \ {J}^\mu_{5}\ =\ \bar\psi \gamma^\mu\gamma_5 \psi \ ,
\end{equation}
and with divergences
\begin{equation}
\partial_\mu {J}^\mu\ =\ 0 \ \ \ ,\ \ \ 
\partial_\mu {J}^\mu_{5}\ =\ 2i \bar\psi\, \mathbb{M}\, \gamma_5 \psi \ -\  N \frac{g^2}{16 \pi^2}\epsilon^{\mu\nu\rho\sigma}\;
tr\left( F_{\mu\nu}F_{\rho\sigma}\right)  \ ,
\end{equation}
where the singlet axial symmetry is of course anomalous. In addition, the Lagrangian is invariant with respect to 
the symmetry transformations
\begin{equation}
\psi\rightarrow e^{-i\theta_\alpha T_\alpha} \psi \ \ \ , \ \ \ \psi\rightarrow e^{-i\theta_\alpha T_\alpha \gamma_5} \psi \ ,
\end{equation}
where the $T_\alpha$  are $SU(N)$ generators (see appendix).
By the standard Noether procedure one defines the associated currents,
\begin{equation}
{J}^\mu_\alpha\ =\ \bar\psi \gamma^\mu T_\alpha \psi \ \ \ , \ \ \ 
{J}^\mu_{5\alpha}\ =\ \bar\psi \gamma^\mu\gamma_5 T_\alpha \psi \ ,
\end{equation}
respectively, with divergences
\begin{equation}
\partial_\mu {J}^\mu_\alpha\ =\ -i \bar\psi\, [\, \mathbb{M}\, ,\, T_\alpha \, ]\, \psi \ \ \ , \ \ \ 
\partial_\mu {J}^\mu_{5\alpha}\ =\ i \bar\psi\, \{ \, \mathbb{M}\, ,\, T_\alpha \, \}\, \gamma_5 \psi \ .
\end{equation}
Therefore, with $N$ degenerate flavors the QCD Lagrangian is $SU(N)_F$ invariant and
in the chiral limit where $\mathbb{M}$ vanishes, there is an
 $SU(N)_R\otimes SU(N)_L$ chiral symmetry generated by the currents
${J}^\mu_{L\alpha} = ({J}^\mu_\alpha - {J}^\mu_{5\alpha})/2$ and ${J}^\mu_{R\alpha} = ({J}^\mu_\alpha + {J}^\mu_{5\alpha})/2$.

The energy-momentum tensor may be written as
\begin{equation}
{T}^{\mu\nu}\ =\ - g^{\mu\nu} {\cal L}_{\rm\scriptstyle QCD} \ -\ 
F^{\mu\rho}_a \, F^{\nu}_{a \, \rho}
\ +\ 
\frac{i}{2}
\bar\psi
\stackrel{{}_\leftrightarrow}{D}{\!}^\mu \gamma^\nu
\psi
\, .
\label{EMtensorQCD}
\end{equation}
From the energy-momentum tensor we can form the Hamiltonian,
\BA
P^0 \ =\ \int\, d^3 \bit{x}\, {T}^{00}\ .
\EA
Here we will assume that chiral symmetry is spontaneously broken through the formation of the condensate
\BA
\mathbb{M}\,\langle\, \Omega \, |\,\bar\psi\psi  \,|\, \Omega\, \rangle\ =\ 
\mathbb{M}\,\langle\, \Omega \, |\,\frac{\partial {\mit T}^{00}}{\partial \mathbb{M}}\,|\, \Omega\, \rangle\ =\
\mathbb{M}\,\frac{\partial {\cal E}_0}{\partial \mathbb{M}}\ \neq\  0 \ ,
\EA
where we have used the Feynman-Hellmann theorem, $|\, \Omega\, \rangle$ represents the (complicated) instant-form QCD vacuum state,
and ${\cal E}_0$ is the QCD vacuum energy. It is straightforward to show that the condensate transforms as 
the $(\bar{\bf N},{\bf N})\oplus({\bf N},\bar{\bf N})$ representation of $SU(N)_R\otimes SU(N)_L$. 
We can compute the vacuum energy in the low-energy effective field theory; i.e. chiral perturbation theory ($\chi$PT)~\cite{Weinberg:1978kz,Gasser:1983yg},
as well. And therefore, 
\BA 
\mathbb{M}\,\frac{\partial {\cal E}_0}{\partial \mathbb{M}}\ =\
\mathbb{M}\,\frac{\partial {\cal E}_0^{\chi{\rm PT}}}{\partial
  \mathbb{M}} \ , 
\EA 
where ${\cal E}_0^{\chi{\rm PT}}$ is the $\chi$PT vacuum energy.  In
the non-linear realization of the chiral group
the Goldstone boson field may be written as
$U(x)=\exp{\left(i\pi_\alpha(x) T_\alpha/F_\pi\right)}$, and the
leading quark mass contribution to the $\chi$PT Lagrangian is
\BA
{\cal L}^{\chi{\rm PT}}_{\rm\scriptstyle QCD} \ =\ v\;tr\left( U \mathbb{M}^\dagger + U^\dagger  \mathbb{M}\right) \ +\ \ldots \ ,
\EA
with $v=M_\pi^2\,F_\pi^2/\mathbb{M}$ and with $M_\pi$ the Goldstone boson mass. One then obtains the Gell-Mann-Oakes-Renner formula~\cite{GellMann:1968rz}.
\BA
-{\mathbb{M}\,\langle\, \Omega \, |\,\bar\psi\psi  \,|\, \Omega\, \rangle}\ =\ \oneht\,N\,M_\pi^2\,F_\pi^2 \ \ +\ \dots \ .
\label{eq:ifgmor}
\EA
It will be a principle goal in what follows to determine what takes the place of this relation in null-plane QCD.

\subsection{Null plane representation}

\noindent The QCD Lagrangian in the null-plane coordinates is obtained by generalizing the results
given in Appendices~\ref{FFdecompose} and \ref{FGdecompose} to the interacting case~\footnote{We follow the notation
and conventions of Ref.~\cite{Belitsky:2005qn}.}. (Note that we work in light-cone gauge, $A^+=0$, throughout.) The 
QCD equations of constraint for the non-dynamical degrees of freedom are
\begin{equation}
\psi_- \ =\ \frac{1}{2 i\! \stackrel{{}_\rightarrow}{\partial^+}}\left( -i \bit{\gamma}^r\!\stackrel{{}_\rightarrow}{D^r}\! \ +\ \mathbb{M} \right) \gamma^+\psi_+
\quad , \quad
\psi^\dagger_- \ =\  \psi^\dagger_+ \gamma^- \left( i \bit{\gamma}^r\!\stackrel{{}_\leftarrow}{D^r}\! \ -\ \mathbb{M} \right) \frac{1}{2 i\! \stackrel{{}_\leftarrow}{\partial^+}}
\end{equation}
for the redundant quark degrees of freedom, and
\BA
\partial^+ A^-_a \ =\ \frac{1}{\partial^+}
D^r_{ab} \partial^+ \bit{A}^r_{b} \ -\ g \frac{1}{\partial^+} \bar\psi_+ \gamma^+ t^a \psi_+ \ ,
\label{eq:freegaugeEOMsol1withI}
\EA
for the redundant gauge degrees of freedom.
The null-plane QCD Lagrangian can then be expressed in terms of the dynamical degrees of freedom as
\begin{eqnarray}
{\tilde {\cal L}}_{\scriptstyle\rm QCD} & = &
i \bar\psi_+ \gamma^+ \partial^- \psi_+
-
\ft{i}{2}
\bar\psi_+ \bit{\gamma}^r \gamma^+ \bit{\gamma}^s
D^r \frac{1}{\partial^+} D^s \psi_+
\nonumber\\
&+&\!\!\!
\ft{i}{2}
\bar\psi_+ \gamma^+ \mathbb{M}^2 \frac{1}{\partial^+} \psi_+
+
\ft{i}{2}
\bar\psi_+ \gamma^+ \mathbb{M}\left( \bit{\gamma}^r g\, t^a\bit{A}^{r}_a\right)\frac{1}{\partial^+}\psi_+
-
\ft{i}{2}
\bar\psi_+ \gamma^+ \mathbb{M}\frac{1}{\partial^+} \left( \bit{\gamma}^r g\, t^a\bit{A}^{r}_a\psi_+\right)
\nonumber\\
&-&\!\!\!
\ft{1}{4} \bit{F}_{a}^{rs} \bit{F}_{a}^{rs}
\ +\ 
\left( \partial^+ \bit{A}^r_{a} \right)
\left( \partial^- \bit{A}^r_{a} \right)
-
\ft12
\left(
\frac{1}{\partial^+}
D^r_{ab} \partial^+ \bit{A}^r_{b}
-
g \frac{1}{\partial^+} \bar\psi_+ \gamma^+ t^a \psi_+
\right)^2
\, .
\end{eqnarray}
The price to pay for working with the physical degrees of freedom in
the null-plane coordinates is a loss of manifest Lorentz covariance, as well as
the appearance of operators which that appear to be non-local in the longitudinal
coordinate.  As in the instant-form, one should view this Lagrangian
as providing a perturbative definition of QCD at large momentum
transfers, where the longitudinal zero modes play no role.
Notice that in null-plane QCD there are two kinds of operators that
depend on the quark-mass matrix~\footnote{To minimize clutter, it will prove convenient to 
define the operator
\begin{eqnarray}
\frac{1}{\partial^+} \left( \bit{\gamma}^r g\, t^a\bit{A}^{r}_a\right)'\psi_+ \ \equiv \
\left( \bit{\gamma}^r g\, t^a\bit{A}^{r}_a\right)\frac{1}{\partial^+}\psi_+
-
\frac{1}{\partial^+} \left( \bit{\gamma}^r g\, t^a\bit{A}^{r}_a\psi_+\right) \ .
\end{eqnarray}
}. One is a kinetic term, quadratic in the quark masses, and the other
is a spin-flip quark-gluon interaction that is linear in the quark masses.

Naturally we expect that null-plane QCD has the same
symmetries as instant-form QCD.  Consider the $U(1)_R\otimes U(1)_L$ transformations,
\begin{equation}
\psi_+\rightarrow e^{-i\theta} \psi_+ \ \ \ , \ \ \ \psi_+\rightarrow e^{-i\theta \gamma_5} \psi_+ \ .
\label{chiraltransu1}
\end{equation}
While baryon number is unaltered in moving to the null-plane
coordinates, this is clearly not the same chiral transformation that
we had in the instant form, as that transformation acts on the
non-dynamical degrees of freedom, $\psi_-$, in a distinct manner and
is therefore complicated on the null-plane. That the chiral symmetry transformations
are different in the two forms of dynamics is essential for what follows. We will return below to
the relation between the chiral symmetries in the instant-form and the
front form, as this will be important in understanding the role of
condensates on the null-plane. The $U(1)_A$ current and its divergence
are~\cite{Wu:2003vn}
\begin{equation}
{\tilde J}^\mu_{5}\ =\ {J}^\mu_{5}  \ -\  i \bar\psi \gamma^\mu \gamma^+\gamma_5\, \mathbb{M}\,  \frac{1}{\partial^+} \psi_+\ ;
\label{eq:HME11qcdu1}
\end{equation}
\begin{equation}
\partial_\mu {\tilde J}^\mu_{5}\ =\ 
\bar\psi_+ \gamma^+\gamma_5 \, \mathbb{M}\,  \frac{1}{\partial^+} \left( \bit{\gamma}^r g\, t^a\bit{A}^{r}_a\right)'\psi_+ 
\ -\  N \frac{g^2}{16 \pi^2}\epsilon^{\mu\nu\rho\sigma}\;
tr\left( F_{\mu\nu}F_{\rho\sigma}\right)  \ .
\end{equation}
Consider the $SU(N)_R\otimes SU(N)_L$ transformations,
\begin{equation}
\psi_+\rightarrow e^{-i\theta_\alpha T_\alpha} \psi_+ \ \ \ , \ \ \ \psi_+\rightarrow e^{-i\theta_\alpha T_\alpha \gamma_5} \psi_+ \ .
\label{chiraltrans}
\end{equation}
The currents associated with eq.~\ref{chiraltrans} are
\begin{equation}
{\tilde J}^\mu_\alpha\ =\  {J}^\mu_\alpha  \ -\  \ft{i}{2} \bar\psi \gamma^\mu \gamma^+ \, [\, \mathbb{M}\, ,\, T_\alpha \, ]\, \frac{1}{\partial^+} \psi_+\ ;
\end{equation}
\begin{equation}
{\tilde J}^\mu_{5\alpha}\ =\ {J}^\mu_{5\alpha}  \ -\  \ft{i}{2} \bar\psi \gamma^\mu \gamma^+\gamma_5 \, \{\, \mathbb{M}\, ,\, T_\alpha \, \}\, \frac{1}{\partial^+} \psi_+ \ ,
\end{equation}
with divergences
\begin{equation}
\partial_\mu {\tilde J}^\mu_\alpha\ =\ \ft{1}{2} \bar\psi \gamma^+ \, [\, \mathbb{M}^2\, ,\, T_\alpha \, ]\, \frac{1}{\partial^+} \psi_+ \ ;
\end{equation}
\begin{equation}
\partial_\mu {\tilde J}^\mu_{5\alpha}\ =\ \ft{1}{2} \bar\psi \gamma^+ \gamma_5 \, [\, \mathbb{M}^2\, ,\, T_\alpha \, ]\, \frac{1}{\partial^+} \psi_+
+\ft{1}{2}
\bar\psi_+ \gamma^+\gamma_5 \{ \, \mathbb{M}\, ,\, T_\alpha \, \}\, \frac{1}{\partial^+} \left( \bit{\gamma}^r g\, t^a\bit{A}^{r}_a\right)'\psi_+ \ .
\end{equation}
For $N$ degenerate flavors, the quark mass matrix is proportional to the identity, the vector current is conserved, and the axial current
and the divergence of the axial current are 
\begin{equation}
{\tilde J}^\mu_{5\alpha}\ =\ {J}^\mu_{5\alpha}  \ -\  i \bar\psi \gamma^\mu \gamma^+\gamma_5\, T_\alpha \, \mathbb{M}\,  \frac{1}{\partial^+} \psi_+\ ;
\label{eq:HME11qcd}
\end{equation}
\begin{equation}
\partial_\mu {\tilde J}^\mu_{5\alpha}\ =\ 
\bar\psi_+ \gamma^+\gamma_5 \, T_\alpha \, \mathbb{M}\,  \frac{1}{\partial^+} \left( \bit{\gamma}^r g\, t^a\bit{A}^{r}_a\right)'\psi_+ \ .
\label{eq:chidivqcd}
\end{equation}
Here note in particular that the null-plane axial-vector current in null-plane QCD evidently takes the form, eq.~\ref{eq:HME11}, expected
on general grounds. 

\subsection{Null-plane charges}

\noindent The null-plane singlet axial charge is defined as
\begin{eqnarray}
{\tilde Q}^5 \ &=& \ \int\, d x^-\, d^2 \bit{x}_\perp\, {\tilde J}^+_{5} \ = \ \int\, d x^-\, d^2 \bit{x}_\perp\,\bar\psi_+\gamma^+ \gamma_5 \, \psi_+  \ ,
\label{eq:QCDu1a}
\end{eqnarray}
where we have used eq.~\ref{eq:HME11qcdu1}. Using the momentum-space representation of $\psi_+$, given in eq.~\ref{eq:freediracmomspace}, one finds
\begin{eqnarray}
{\tilde Q}^5  &=& \sum_{\lambda = \uparrow\downarrow}2\lambda
\int \frac{d k^+ d^2 \bit{k}_\perp}{2 k^+ (2 \pi)^3}
 \left\{  b^\dagger_\lambda (k^+ , \bit{k}_\perp) b_\lambda (k^+ , \bit{k}_\perp)   +
d^\dagger_\lambda (k^+ , \bit{k}_\perp) d_\lambda (k^+ , \bit{k}_\perp) \right\}.
\end{eqnarray}
Comparison with eq.~\ref{eq:freehelquark} one sees that the singlet axial charge coincides (up to a factor of two) with the free-fermion helicity operator.
This of course explains why the quark mass term in the free-fermion theory is a chiral invariant; on the null-plane, chiral symmetry breaking
in the free-fermion theory implies breaking of rotational invariance in the transverse plane.

Similarly, the null-plane non-singlet vector and chiral charges are, respectively, 
\begin{eqnarray}
{\tilde Q}_\alpha \ &=& \ \int\, d x^-\, d^2 \bit{x}_\perp\, {\tilde J}^+_{\alpha} \ = \ \int\, d x^-\, d^2 \bit{x}_\perp\,\bar\psi_+\gamma^+ \, T_\alpha \psi_+  \ ; \\
{\tilde Q}^5_\alpha \ &=& \ \int\, d x^-\, d^2 \bit{x}_\perp\, {\tilde J}^+_{5\alpha} \ = \ \int\, d x^-\, d^2 \bit{x}_\perp\,\bar\psi_+\gamma^+ \gamma_5 \, T_\alpha \psi_+  \ ,
\label{eq:QCDcc}
\end{eqnarray}
and using the momentum-space representation of $\psi_+$, given in eq.~\ref{eq:freediracmomspace}, one finds
\begin{eqnarray}
\hskip-1.9em{\tilde Q}_\alpha  &=& \sum_{\lambda = \uparrow\downarrow}
\int \frac{d k^+ d^2 \bit{k}_\perp}{2 k^+ (2 \pi)^3}
 \left\{  b^\dagger_\lambda (k^+ , \bit{k}_\perp) T_\alpha b_\lambda (k^+ , \bit{k}_\perp)   -
d^\dagger_\lambda (k^+ , \bit{k}_\perp) T^T_\alpha d_\lambda (k^+ , \bit{k}_\perp) \right\} \ ; \\
\hskip-1.9em{\tilde Q}^5_\alpha  &=& \sum_{\lambda = \uparrow\downarrow}2\lambda
\int \frac{d k^+ d^2 \bit{k}_\perp}{2 k^+ (2 \pi)^3}
 \left\{  b^\dagger_\lambda (k^+ , \bit{k}_\perp) T_\alpha b_\lambda (k^+ , \bit{k}_\perp)   +
d^\dagger_\lambda (k^+ , \bit{k}_\perp) T^T_\alpha d_\lambda (k^+ , \bit{k}_\perp) \right\}.
\end{eqnarray}
One readily checks that the null-plane chiral algebra, eqs.~\ref{eq:LCalga} and \ref{eq:LCalgb}, is satisfied by these charges. 
As these charge are written as sums of number operators that count the number
of quarks and anti-quarks, both chiral charges annihilate the vacuum, and we have
\BA
{\tilde Q}^\alpha\, |\, 0\,\rangle\, =\, {\tilde Q}_5^\alpha\, |\, 0\,\rangle\, =\, 0 \ ,
\label{eq:vacuumann2}
\EA
as expected on the general grounds presented above. One then has
\BA
[\, {\tilde Q}^\alpha\, ,\, \psi_+ \, ]\, =\,  -T^\alpha\,\psi_+ \ \ ; \ \ 
[\, {\tilde Q}^\alpha_5\, ,\, \psi_+ \, ]\, =\,  -\gamma_5\,T^\alpha\,\psi_+ \ .
\EA
Breaking down the fields into left- and right-handed components, 
\BA
\psi_{+R} \ =\ \ft{1}{2} (1+\gamma_5) \psi_+ \qquad , \qquad 
\psi_{+L} \ =\ \ft{1}{2} (1-\gamma_5) \psi_+ 
\EA
and, using the results of Appendix~\ref{FFdecompose}, one verifies the fermion transformation properties with respect to $SU(N)_R\otimes SU(N)_L$:
\BA
\psi_{+R}\ =\ \psi_{+\uparrow}\ &\in& \ ({\bf 1},{\bf N}) \qquad , \qquad \psi_{+R}^\dagger \ =\ \psi_{+\downarrow}^\dagger\ \in \ ({\bf 1},\bar{\bf N})\ ; \\
\psi_{+L}\ =\ \psi_{+\downarrow}\ &\in& \ ({\bf N},{\bf 1}) \qquad , \qquad \psi_{+L}^\dagger \ =\ \psi_{+\uparrow}^\dagger\ \ \in \ (\bar{\bf N},{\bf 1})\ ,
\label{eq:chiralassign}
\EA
and the helicity eigen-equations of the quarks
\BA
\Sigma_{12}\;\psi_{+\uparrow}& =& {\textstyle\frac{1}{2}}\psi_{+\uparrow} \ ;\\
\Sigma_{12}\;\psi_{+\downarrow}& =& -{\textstyle\frac{1}{2}}\psi_{+\uparrow} \ ,
\EA
where the helicity operator, $\Sigma_{12}$, is defined in Appendix~\ref{FFdecompose}.

\subsection{Chiral symmetry breaking Hamiltonians}

\noindent Using the results of the previous section, it is
straightforward to find the transformation properties of the
symmetry-breaking parts of the reduced Hamiltonians.  Define the
operators~\footnote{From here forward we will use the definition:
\begin{equation}
\frac{1}{\partial_{\mathbb{M}}^+} \equiv \mathbb{M}\,\frac{1}{\partial^+} \ .\nonumber
\end{equation}
}:
\begin{equation}
{\tilde D}_{5\alpha} \ \equiv \ \partial_\mu {\tilde J}^\mu_{5\alpha}\ =\ 
\bar\psi_+ \gamma^+\gamma_5 \, T_\alpha \, \frac{1}{\partial_\mathbb{M}^+} \left( \bit{\gamma}^r g\, t^a\bit{A}^{r}_a\right)'\psi_+ \ ;
\label{eq:axcurr2}
\end{equation}
\begin{equation}
{\tilde D}_{5} \ \equiv \ \bar\psi_+ \gamma^+\gamma_5 \, \frac{1}{\partial_\mathbb{M}^+} \left( \bit{\gamma}^r g\, t^a\bit{A}^{r}_a\right)'\psi_+ \ ;
\end{equation}
\begin{equation}
{\tilde D}_{\alpha} \ \equiv \ \bar\psi_+ \gamma^+\, T_\alpha \, \frac{1}{\partial_\mathbb{M}^+} \left( \bit{\gamma}^r g\, t^a\bit{A}^{r}_a\right)'\psi_+ \ ;
\end{equation}
\begin{equation}
{\tilde D} \ \equiv \ \bar\psi_+ \gamma^+\, \frac{1}{\partial_\mathbb{M}^+} \left( \bit{\gamma}^r g\, t^a\bit{A}^{r}_a\right)'\psi_+ \ .
\end{equation}
It is then a textbook exercise to find
\BA
&&\hfill[\, {\tilde Q}^\alpha_5\, ,\, {\tilde D}^\beta_5 \, ] \, =\,  \frac{1}{N}\,\delta^{\alpha\beta}\;{\tilde D} \ +\ d^{\alpha\beta\gamma}\,{\tilde D}^\gamma \ ; 
\label{eq:algsola}\\
&&\hfill[\, {\tilde Q}^\alpha_5\, ,\, {\tilde D}\, ] \, =\,  2\;{\tilde D}_5^\alpha \ ; 
\label{eq:algsolb}\\
&&[\, {\tilde Q}^\alpha_5\, ,\, {\tilde D}^\beta \, ]\, =\,  \frac{1}{N}\,\delta^{\alpha\beta}\;{\tilde D}_5 \ +\ d^{\alpha\beta\gamma}\,{\tilde D}_5^\gamma \ ; 
\label{eq:algsolc}\\
&&[\, {\tilde Q}^\alpha_5\, ,\, {\tilde D}_5\, ]\, =\,  2\;{\tilde D}^\alpha \ .
\label{eq:algsold}
\EA
It follows that the $2N^2$ operators $({\tilde D}_{5\alpha},{\tilde D}_{5},{\tilde D}_{\alpha},{\tilde D})$ fill out the
$(\bar{\bf N},{\bf N})\oplus({\bf N},\bar{\bf N})$ representation of $SU(N)_R\otimes SU(N)_L$. 

The null-plane Hamiltonian $P^-$ is:
\BA
P^- \ = \ \int\, d x^-\, d^2 \bit{x}_\perp\, {T}^{-+}\  ,
\EA
and therefore the chiral-symmetry breaking part of this Hamiltonian is given by:
\BA
P_{({\bf N},{\bf N})}^- \ \equiv \ -\ft{i}{2}\,\int\, d x^-\, d^2 \bit{x}_\perp\, {\tilde D} \ .
\EA
One readily checks that this is consistent with eqs.~\ref{eq:mixedchiralc} and \ref{eq:axcurr2}.

One then finds the symmetry breaking parts of the reduced QCD Hamiltonians:
\BA
M^2_{({\bf N},{\bf N})} & = & -i P^+ \,
\int\, d x^-\, d^2 \bit{x}_\perp\, \bar\psi_+ \gamma^+\, \frac{1}{\partial_\mathbb{M}^+} \left( \bit{\gamma}^r g\, t^a\bit{A}^{r}_a\right)'\psi_+  \ ; 
\label{eq:symmbreakhamsa}\\
\left(M{\cal J}_r\right)_{({\bf N},{\bf N})} & = &   i{\textstyle{\frac{1}{2}}}\epsilon_{rs}\,P^+
{\int}\, d x^-\, d^2 \bit{x}_\perp\, {\Gamma}_s\, \bar\psi_+ \gamma^+\, \frac{1}{\partial_\mathbb{M}^+} \left( \bit{\gamma}^r g\, t^a\bit{A}^{r}_a\right)'\psi_+  \ ,
\label{eq:symmbreakhamsb}
\EA
where, in addition, we have used eqs.~\ref{eq:mixedchirald} and \ref{eq:chidivqcd} to obtain the reduced Hamiltonian for spin.
All chiral symmetry breaking in null-plane QCD is contained in these two operators. 

Using eqs.~\ref{eq:mixedchiralc},  \ref{eq:mixedchirald} and \ref{eq:algsola} one finds
\BA
{}[{\tilde Q}^\beta_5\, ,\, [\, {\tilde Q}^\alpha_5\, ,\, M^2\, ]] &=& -2iP^+\;\int\, d x^-\, d^2 \bit{x}_\perp\;\left(\;
\frac{1}{N}\,\delta^{\alpha\beta}\,{\tilde D} \ +\ d^{\alpha\beta\gamma}\,{\tilde D}^\gamma\; \right) \ ; \label{eq:fundcona}\\
{}[{\tilde Q}^\beta_5\, ,\, [\, {\tilde Q}^\alpha_5\, ,\, M{\cal J}_r\, ]] &=& i\epsilon_{rs}P^+\;\int\, d x^-\, d^2 \bit{x}_\perp\,\Gamma_s\;\left(\;
\frac{1}{N}\,\delta^{\alpha\beta}\,{\tilde D} \ +\ d^{\alpha\beta\gamma}\,{\tilde D}^\gamma\; \right) \ .
\label{eq:fundconb}
\EA
Acting on these equations with $\delta^{\alpha\beta}$ and $d^{\alpha\beta\gamma}$, and using the identities in Appendix~\ref{suN} gives
\BA
&&-2iP^+ \int\, d x^-\, d^2 \bit{x}_\perp\, {\tilde D} \ =\ \frac{N}{N^2-1}\, [{\tilde Q}^\alpha_5\, ,\, [\, {\tilde Q}^\alpha_5\, ,\, M^2\, ]]\ ; \\
&&-2iP^+ \int\, d x^-\, d^2 \bit{x}_\perp\, {\tilde D}_\gamma \ =\ d_{\alpha\beta\gamma}\frac{N}{N^2-4}\, [{\tilde Q}^\beta_5\, ,\, [\, {\tilde Q}^\alpha_5\, ,\, M^2\, ]]\ .
\EA
Therefore, eq.~\ref{eq:fundcona} can be written as
\begin{eqnarray}
\hskip-1.4em[{\tilde Q}^\beta_5 , [ {\tilde Q}^\alpha_5 , M^2]]=\frac{1}{N^2-1}\delta^{\alpha\beta}[{\tilde Q}^\gamma_5 , [ {\tilde Q}^\gamma_5 , M^2 ]]
+ \frac{N}{N^2-4}d^{\alpha\beta\gamma}d^{\mu\nu\gamma}[ {\tilde Q}^\mu_5 , [ {\tilde Q}^\nu_5 , M^2 ]]\ ,
\end{eqnarray} 
and eq.~\ref{eq:fundconb} takes the same form but with $M^2$ replaced by $M {\cal J}_\pm$.  Defining the projection operator
\BA {\cal P}^{\alpha\beta ; \mu\nu}\ \equiv\
\delta^{\alpha\nu}\delta^{\beta\mu}\ -\
\frac{1}{N^2-1}\;\delta^{\alpha\beta}\delta^{\mu\nu} \ -\
\frac{N}{N^2-4}\;d^{\alpha\beta\gamma}d^{\mu\nu\gamma} \ , 
\EA 
we can express the constraints on the reduced Hamiltonians in compact notation as:
\BA
{\cal P}^{\alpha\beta ;\mu\nu}\;[{\tilde Q}^\mu_5\, ,\, [\, {\tilde Q}^\nu_5\, ,\, M^2]]\ =\ {\cal P}^{\alpha\beta ;\mu\nu}\;[{\tilde Q}^\mu_5\, ,\, [\, {\tilde Q}^\nu_5\, ,\, M {\cal J}_\pm]]\ =\ 0 \ .
\label{eq:fundmix}
\EA 
These are quite possibly the most important equations in null-plane
QCD, as they are the mathematical expression of the specific way in
which the internal symmetries and Poincar\'e symmetries intersect.
These equations were obtained originally in
Refs.~\cite{Weinberg:1969hw,Weinberg:1969db,Weinberg:1990xn} by
considering the most general form of Goldstone-boson-hadron scattering
amplitudes in specially-designed Lorentz frames, and using input from Regge-pole
theory expectations of their high-energy behavior.  Note that the
projection operator, ${\cal P}^{\alpha\beta ; \mu\nu}$, has four
adjoint indices and is, as shown in Ref.~\cite{Weinberg:1969hw}
related to the interactions of Goldstone bosons (in the t-channel of
Goldstone-boson-hadron scattering), which are in the adjoint of
$SU(N)_F$ and whose scattering amplitudes therefore transform as the
product of two adjoints. In the case of two flavors, where ${\bf
  3}\otimes{\bf 3}={\bf 1}\oplus{\bf 3}\oplus{\bf 5}$, it projects out
the ${\bf 5}$-dimensional representation ($I=2$) and in the case of
three flavors, where ${\bf 8}\otimes{\bf 8}={\bf 1}\oplus{\bf
  8}\oplus{\bf 8}\oplus{\bf 10}\oplus\bar{\bf 10}\oplus{\bf 27}$, it
projects out the ${\bf 10}$, $\bar{\bf 10}$, and ${\bf
  27}$-dimensional representations.  As shown above, these are the
representations that cannot be formed from a single quark bilinear;
i.e. they are not contained in $(\bar{\bf N},{\bf N})\oplus({\bf
  N},\bar{\bf N})$, as is clear from direct inspection of
eqs.~\ref{eq:fundcona} and \ref{eq:fundconb}.

\subsection{Gell-Mann-Oakes-Renner relation recovered}

\noindent We are now in a position to address the fate of instant-form QCD chiral-symmetry breaking condensates
in null-plane QCD. Again using the Feynman-Hellmann theorem we find
\BA 
\mathbb{M}\,\langle\, 0 \, |\,\frac{\partial T^{-+}}{\partial \mathbb{M}}\,|\, 0\, \rangle &=& \mathbb{M}\,\frac{\partial {\tilde {\cal E}}_0}{\partial
  \mathbb{M}}\ =\ 
\mathbb{M}\,\frac{\partial {\cal E}_0}{\partial \mathbb{M}}\ =\
\mathbb{M}\,\frac{\partial {\cal E}_0^{\chi{\rm PT}}}{\partial
  \mathbb{M}} \ , 
\EA 
where $|\, 0\, \rangle$ represents the null-plane QCD vacuum
state, and ${\tilde {\cal E}}_0$ is the null-plane QCD vacuum energy. In this equation we have also expressed
that physics is independent of the choice of coordinates. Therefore calculation
of the leading quark-mass contribution to the vacuum energy must be independent of the quantization surface,
and should be the same whether one works with the fundamental degrees of freedom, or with the Goldstone bosons
in the infrared. One then has
\BA 
\mathbb{M}\,\frac{\partial {\tilde {\cal E}}_0}{\partial \mathbb{M}}\ =\ -\mathbb{M}\,\langle\, 0\, |\,i\;\bar\psi_+ \gamma^+
\frac{1}{\partial^+_\mathbb{M}} \psi_+\,|\, 0\, \rangle  +
\langle\, 0\, |\,  \ft{i}{2}
\bar\psi_+ \gamma^+ \frac{1}{\partial^+_\mathbb{M}} \left( \bit{\gamma}^r g\, t^a\bit{A}^{r}_a\right)'\psi_+ 
\,|\, 0\, \rangle \ .
\EA 
The second term must vanish as the chiral charges annihilate the vacuum and therefore there can be no
chiral-symmetry breaking condensates. Operationally one sees this directly by 
taking the vacuum expectation value of eq.~\ref{eq:algsola} which gives
\BA 
\langle\, 0\, |\,  \ft{i}{2}
\bar\psi_+ \gamma^+ \frac{1}{\partial^+_\mathbb{M}} \left( \bit{\gamma}^r g\, t^a\bit{A}^{r}_a\right)'\psi_+ 
\,|\, 0\, \rangle \ =\ 0.
\EA 
We are then left with the null-plane expression of the Gell-Mann-Oakes-Renner relation:
\BA {\mathbb{M}\,\langle\, 0\, |\,i\;\bar\psi_+ \gamma^+
  \frac{1}{\partial^+_\mathbb{M}} \psi_+\,|\, 0\, \rangle}\ =\ \oneht\,
N\,M_\pi^2\,F_\pi^2 \ \ +\ \dots \ .  
\label{eq:npgoar}
\EA Hence, a chiral-symmetry breaking condensate in the instant-form
formulation of QCD has been replaced by a chiral-symmetry conserving
condensate in the null-plane formulation. Note that while the operator
naively vanishes in the chiral limit, the matrix element is infrared
singular and therefore it need not, and indeed cannot, vanish in the
chiral limit~\footnote{This expression of the Gell-Mann-Oakes-Renner
  formula was found previously in Ref.~\cite{Wu:2003vn} using the
  methods that will be described below.}.  It would be very
interesting to define the relevant operator non-perturbatively and
calculate this condensate directly, perhaps using transverse lattice
gauge theory
methods~\cite{Bardeen:1979xx,Burkardt:1998ws,Dalley:1998bj,Dalley:2001gj,Burkardt:2001mf,Burkardt:2001dy,Burkardt:2001jg,Dalley:2002nj,Bratt:2004wq}. Note
that {\it a priori} knowledge of the singlet condensate in
eq.~\ref{eq:npgoar} is not very different to {\it a priori} knowledge of the symmetry-breaking quark condensate in
eq.~\ref{eq:ifgmor}. In both cases, it is necessary to keep the quark masses finite and only
at the very end take the chiral limit~\cite{Banks:1979yr}.

\subsection{Condensates on a null-plane}

\noindent We will now derive the Gell-Mann-Oakes-Renner relation in a different way
which will suggest a general prescription for expressing all
instant-form condensates with null-plane condensates.
While the left- and right-handed components of $\psi_+$ transform irreducibly with respect
to the null-plane chiral charges, the transformation properties of $\psi$ are complicated
by the presence of the non-dynamical component $\psi_-$. Indeed one finds
\BA [\, {\tilde Q}^\alpha_5\, ,\, \psi \, ]\, =\,
-\gamma_5\,T^\alpha\,\psi \ -\
i\;\gamma_5\;\gamma^+\;T^\alpha\;\frac{1}{\partial^+_\mathbb{M}} \psi \ ,  
\label{eq:psitrans}
\EA 
from which it follows that 
\BA
\psi_{R}\, ,\, \psi_{L}\ \in \ ({\bf 1},{\bf N})\oplus ({\bf N},{\bf 1}) \quad , \quad
\psi_{R}^\dagger\, ,\,\psi_{L}^\dagger \ \in \ ({\bf 1},\bar{\bf
  N})\oplus(\bar{\bf N},{\bf 1}) \ .  \EA 
Since the left- and right-handed components of the quark field
transform reducibly with respect to the chiral group, generally
products of bilinear operators of the form $\bar\psi\Gamma\psi$ will have
complicated reducible chiral transformation properties. However, QCD
operators built out of these bilinears will always have a component that
transforms as a chiral singlet. We will now see, for the simplest
example, that this is essential to the consistency of the null-plane
formulation. Consider the transformation properties of the following
set of bilinears:
\BA
{D}^\alpha_{5} \ &\equiv& \ \bar\psi\,\gamma_5\,T^\alpha\,\psi\ \ , \ \
{D}_{5} \ \equiv \ \bar\psi\,\gamma_5\,\psi\ ; \\
{D}^{\alpha} \ &\equiv& \ \bar\psi\,T^\alpha\,\psi\ \ \ , \ \ \ \
{D} \ \equiv \ \bar\psi\,\psi\ .
\EA
Is is again simple to check that these operators fill out the
$(\bar{\bf N},{\bf N})\oplus({\bf N},\bar{\bf N})$ representation of $SU(N)_R\otimes SU(N)_L$
with respect to the instant-form chiral charges $Q_5^\alpha$.
Now consider the transformation properties of these operators 
with respect to the null-plane chiral charges. One finds
\BA 
&&[\, {\tilde Q}^\alpha_5\, ,\,{D}^\beta_{5} \, ]\,
=\, -\frac{1}{N}\,\delta^{\alpha\beta}\,\left(\,D + \
  i\;
  \bar\psi_+ \gamma^+  \frac{1}{\partial^+_\mathbb{M}} \psi_+\right) \label{eq:GORf} \nonumber \\
&&\quad\qquad\qquad\qquad-\,d^{\alpha\beta\gamma}\left(\,D^\gamma \ +\
  i\;\bar\psi_+ \gamma^+ T^\gamma \frac{1}{\partial^+_\mathbb{M}}
  \psi_+\right) \ ; \\
&&[\, {\tilde Q}^\alpha_5\, ,\,{D} \, ]\,
=\, -2\,D^\alpha_5 \ ; \\
&&[\, {\tilde Q}^\alpha_5\, ,\,{D}^\beta \, ]\,
=\, -\frac{1}{N}\,\delta^{\alpha\beta}\,D_5 \ -\ d^{\alpha\beta\gamma}\,D_5^\gamma\nonumber\\
&&\quad\qquad\qquad\qquad
  +\,f^{\alpha\beta\gamma}\,\bar\psi_+ \gamma^+ \gamma_5 T^\gamma \frac{1}{\partial^+_\mathbb{M}}
  \psi_+\ ; \\
&&[\, {\tilde Q}^\alpha_5\, ,\,{D}_5 \, ]\,
=\, -2\,D^\alpha\ -\ 2i  \bar\psi_+ \gamma^+ T^\alpha \frac{1}{\partial^+_\mathbb{M}} \psi_+ \ .
\EA 
To close the algebra we must add, in addition, the commutation relations:
\BA 
&&[\, {\tilde Q}^\alpha_5\, ,\,\bar\psi_+ \gamma^+  \frac{1}{\partial^+_\mathbb{M}} \psi_+ \, ]\ =\ 0 \ ; \\
&&[\, {\tilde Q}^\alpha_5\, ,\,\bar\psi_+ \gamma^+ T^\beta \frac{1}{\partial^+_\mathbb{M}} \psi_+ \, ]\ =\ 
i\,f^{\alpha\beta\gamma} \, \bar\psi_+ \gamma^+ \gamma_5 T^\gamma \frac{1}{\partial^+_\mathbb{M}} \psi_+ \ ; \\
&&[\, {\tilde Q}^\alpha_5\, ,\,\bar\psi_+ \gamma^+ \gamma_5 T^\beta \frac{1}{\partial^+_\mathbb{M}} \psi_+ \, ]\ =\ 
i\,f^{\alpha\beta\gamma} \, \bar\psi_+ \gamma^+ T^\gamma \frac{1}{\partial^+_\mathbb{M}} \psi_+ \ .
\EA
Hence the full set of operators transform as the reducible $4N^2$-dimensional
$({\bf 1},{\bf 1})\oplus({\bf 1},{\bf {\cal A}})\oplus({\bf {\cal A}},{\bf 1})\oplus (\bar{\bf N},{\bf  N})\oplus({\bf N},\bar{\bf N})$ 
representation of $SU(N)_R\otimes SU(N)_L$, where here ${\bf {\cal A}}$ denotes the $SU(N)$ adjoint
representation. In particular one see that
\BA
\bar\psi\psi \in \ (\bar{\bf N},{\bf N})\oplus({\bf N},\bar{\bf N})\oplus({\bf 1},{\bf 1})\oplus\ldots \ ,
\EA 
and therefore transforms reducibly. This is verified by direct calculation which gives~\footnote{Note that the second term, which is breaks chiral symmetry
and is independent of the interaction does not appear in the free fermion Lagrangian as it is cancelled by a piece coming from the other kinetic term, as must be the
case in order that the Lagrangian commute with the helicity operator. }
\BA
\hskip-1.3em\mathbb{M}\,\bar\psi\psi \ =\ -i\,\mathbb{M}\,\bar\psi_+ \gamma^+ \frac{1}{\partial^+_\mathbb{M}} \psi_+ \ +\  
\ft{i}{2}\bar\psi_+ \gamma^+\, \frac{1}{\partial_\mathbb{M}^+} \left( \bit{\gamma}^r g\, t^a\bit{A}^{r}_a\right)'\psi_+\ .
\label{eq:psibarpsiexpl}
\EA 
Taking the vacuum expectation value of eq.~\ref{eq:psibarpsiexpl} (or eq.~\ref{eq:GORf}) gives the general
solution~\cite{Wu:2003vn}
\BA
\langle\, 0\, |\,\bar\psi\psi\,|\, 0\, \rangle\ = \ -\langle\, 0\, |\,i\;\bar\psi_+ \gamma^+  \frac{1}{\partial^+_\mathbb{M}} \psi_+\,|\, 0\, \rangle \ .
\label{eq:WuZhang}
\EA 
Therefore only the singlet part of $\bar\psi\psi$ can acquire a vacuum
expectation value on the null plane, as must be the case since
$SU(N)_R\otimes SU(N)_L$ is a symmetry of the null-plane vacuum
state. This argument readily generalizes to any chiral symmetry
breaking Lorentz scalar operator, ${\mathcal O}$, that one can build
out of products of quark bilinears in instant-form QCD. One can write
\BA
{\mathcal O} \ =\  \sum_{\bf{\mathcal R}}\,{\mathcal O}_{\bf{\mathcal R}} \ =\  \sum_{\tilde{\bf{\mathcal R}}}\,{\cal O}_{\tilde{\bf{\mathcal R}}}\ + \ {\mathcal O}_{\tilde{\bf 1}} 
\label{eq:gencondensate}
\EA 
where $\bf{\mathcal R}$ is a non-trivial chiral representation with respect to the instant-form chiral charges, $Q_{5\alpha}$,
and $\tilde{\bf{\mathcal R}}$ (${\tilde{\bf 1}}$) is a non-trivial (the singlet) representation with respect to the 
front-form chiral charges, ${\tilde Q}_{5\alpha}$. Unless protected by another symmetry, ${\mathcal O}$ has a non-vanishing 
vacuum expectation value, which can be expressed as
\BA
\langle\, \Omega \, |\,{\mathcal O} \,|\, \Omega\, \rangle\ =\ 
\langle\, \Omega \, |\,\sum_{\bf{\mathcal R}}\,{\mathcal O}_{\bf{\mathcal R}}\,|\, \Omega\, \rangle\ =\ 
\langle\, 0 \, |\,{\mathcal O}_{\tilde{\bf 1}} \,|\, 0\, \rangle \ \neq \ 0 \ .
\label{eq:gencond}
\EA
Note that the final equality expresses an equivalence between a matrix element evaluated
in the instant form and one in the front form. This equality ensures that physics is
unmodified in moving between the two forms of dynamics.
Therefore all instant-form chiral symmetry breaking QCD condensates map to chiral
symmetry conserving condensates in the front-form. The presence of the singlet
part of the operator can always be traced to the reducible chiral transformation
property of $\psi$ given in eq.~\ref{eq:psitrans}. For the case at hand, with 
${\mathcal O}=\bar\psi\psi$, we have
\BA
\langle\, \Omega\, |\,\bar\psi\psi\,|\, \Omega\, \rangle \ = \ \langle\, 0\, |\,\bar\psi\psi\,|\, 0\, \rangle \ ,
\label{eq:psibarpsicond}
\EA 
which together with eq.~\ref{eq:WuZhang}, provides the desired link between the instant-form and front-form expressions of
the Gell-Mann-Oakes-Renner relation.

The general relation, eq.~\ref{eq:gencond} is important for the
consistency of null-plane QCD, as it demonstrates that, as expected,
the QCD vacuum energy is unaltered in moving from the instant-form to
the front-form description, and these relations must, of course, exist
in order that the operator product expansion be independent of the
choice of quantization surface.  We see that a symmetry-breaking
condensate can form in the instant-form coordinates with an asymmetric
vacuum which is equal to a corresponding symmetry-preserving
condensate in the null-plane description with a symmetric vacuum. The
condensate relation eq.~\ref{eq:psibarpsicond} is one of an infinite
number of relations which translates condensates which break chiral
symmetry in the instant form to null-plane condensates which transform
as chiral singlets.

\section{Consequences of the operator algebra}
\label{sec:consequences}

\subsection{Summary of the null-plane QCD description}

\noindent Before considering the consequences of the null-plane QCD
operator algebra, we will summarize the picture of chiral symmetry
breaking that we have so far established. While the null-plane QCD
vacuum state is chirally invariant, chiral symmetry is spontaneously
broken by the three reduced Hamiltonians that have contributions,
$M^2_{({\bf N},{\bf N})}$ and $\left(M{\cal J}_r\right)_{({\bf N},{\bf
    N})}$, which transform as $(\bar{\bf N},{\bf N})\oplus({\bf
  N},\bar{\bf N})$ with respect to $SU(N)_R\otimes SU(N)_L$.  The
three reduced Hamiltonians satisfy the constraints,
eq.~\ref{eq:fundmix}.  In addition to these signatures of chiral
symmetry breaking, the three reduced Hamiltonians, together with the
generator of rotations on the transverse plane together generate the
$U(2)$ dynamical sub-group of the null-plane Poincar\'e algebra,
eq.~\ref{eq:dynalgarbB}.  And finally, the null-plane vector and
chiral charges satisfy the $SU(N)_R\otimes SU(N)_L$ algebra,
eqs.~\ref{eq:LCalga} and \ref{eq:LCalgb}.  The entire set of
Lie-brackets provide all of the constraints that exist among the
generators of the internal and space-time symmetries in null-plane
QCD.  The consequences of chiral symmetry breaking for the spectrum
and spin of QCD are contained in the symmetry-breaking parts of the
reduced Hamiltonians.

\subsection{Recovery of spin-flavor symmetries}

\noindent In searching for solutions of the algebraic system that
mixes the chiral charges and the reduced Hamiltonians, one may worry
about the existence of no-go theorems that forbid non-trivial algebras
that mix space-time and internal symmetries. In the null-plane
formulation the no-go theorems are avoided because it is only the
dynamical part, ${\cal D}$, of the null-plane Poincar\'e algebra that
mixes with the internal symmetry generators~\cite{Leutwyler:1977vz}.
Unfortunately, a direct general solution of the null-plane QCD
operator algebra in the general case appears difficult. However, there
is a limiting case in which the algebra yields an important
non-trivial solution.  Here we will treat the QCD operator algebra as
an abstract operator algebra and consider the limit in which the
chiral-symmetry breaking part of the reduced Hamiltonian $M^2$ can be
treated as a perturbation. However, one should keep in mind that
matrix elements of the operator relations between hadronic states
must eventually be taken in order to extract observables.
We first define
\BA
[\, {\tilde Q}^\alpha_5\, ,\, M\, ] \, \equiv \,  \epsilon^\alpha \ ,
\EA
and neglect terms of ${\cal O}(\epsilon)$. This implies that all
chiral symmetry breaking occurs in the spin Hamiltonians.
This limit is non-trivial, as we have shown above in 
section~\ref{susec:spin} that the spin Hamiltonians alone imply the presence
of Goldstone bosons. In this limit, the QCD operator algebra reduces to
\BA
[\, {\cal J}_i \, ,\,   {\cal J}_j  \, ] \, =\, i\,\epsilon_{ijk}\,  {\cal J}_k
\EA
which generates $SU(2)$ spin, and the $SU(N)_R\otimes SU(N)_L$ algebra,
\BA
[\, {\tilde Q}^\alpha\,  ,\, {\tilde Q}^\beta\, ]\, =\, i\,f^{\alpha\beta\gamma} \, {\tilde Q}^\gamma \ \ , \ \ 
[\, {\tilde Q}^\alpha_5\, ,\, {\tilde Q}^\beta\, ]\, =\,  i\,f^{\alpha\beta\gamma}\,  {\tilde Q}^\gamma_5 \ \ , \ \
[\, {\tilde Q}^\alpha_5\, ,\, {\tilde Q}^\beta_5\, ] \, = \,  i\,f^{\alpha\beta\gamma} \, {\tilde Q}^\gamma  \ .
\label{eq:QCDoa}
\EA
The remaining non-trivial mixed commutator is for the spin Hamiltonian:
\BA 
{\cal P}^{\alpha\beta ;\mu\nu}\;[{\tilde Q}^\mu_5\, ,\, [\, {\tilde
  Q}^\nu_5\, ,\, {\cal J}_\pm]]\ =\ 0 \ .  
\label{eq:OPQcomm2b}
\EA 
Now this simplified algebra can be put into a more familiar form. Consider
an operator $G_{\alpha i}$ which transforms in the adjoint of  $SU(N)$ and as a rotational
vector in the sense that
\begin{eqnarray}
&&[\, {\cal J}_i \, ,\,   {G}_{\alpha j}  \, ] \ = \ i\,\epsilon_{ijk}\,  {G}_{\alpha k} \ ; \\
&&[\, {\tilde Q}_\alpha \, ,\,   {G}_{\beta i}  \, ]  \ = \ i\,f_{\alpha\beta\gamma}\,  {G}_{\gamma i} \ .
\end{eqnarray}
In general, the commutator of $G^{\alpha i}$ with itself may be expressed as
\BA
[\, G_{\alpha i} \, ,\, G_{\beta j} \, ] \, =\, 
i\,f_{\alpha\beta\gamma}\, {\cal A}_{ij,\gamma}\, +\, i\,\epsilon_{ijk}\,  {\cal B}_{\alpha\beta,k} \ ,
\label{eq:su4commgen}
\EA
where ${\cal A}_{ij,\gamma}={\cal A}_{ji,\gamma}$ and ${\cal
  B}_{\alpha\beta,k}={\cal B}_{\beta\alpha,k}$.  Now we identify
$G^{\alpha 3}\ \equiv \ {\tilde Q}^\alpha_5$. From
eq.~\ref{eq:QCDoa} it then follows that ${\cal
  A}_{33,\alpha}={\tilde Q}_\alpha$. Rotational invariance then
implies ${\cal A}_{ij,\alpha}=\delta_{ij}{\tilde Q}_\alpha$.  By
considering Jacobi identities of ${\cal J}_i$ and ${\tilde Q}_\alpha$
with the commutator in eq.~\ref{eq:su4commgen} one finds, respectively,
\BA
&&[\, {\tilde Q}_\gamma \, ,\,   {\cal B}_{\alpha\beta,i}  \, ] \, =\, i\,f_{\gamma\beta\mu}\, {\cal B}_{\alpha\mu,i} \, +\, i\,f_{\gamma\alpha\mu}\, {\cal B}_{\beta\mu,i}\ ; \\
&&[\, {\cal J}_i \, ,\,   {\cal B}_{\alpha\beta,j}  \, ] \, =\, i\,\epsilon_{ijk}\, {\cal B}_{\alpha\beta,k}\ ,
\EA
which simply indicate that ${\cal B}_{\alpha\beta,i}$ transforms as a rank-two $SU(N)$ tensor and a
rotational vector.

To obtain ${\cal B}_{\alpha\beta,i}$ we use eq.~\ref{eq:OPQcomm2b} to find:
\BA
{\cal P}_{\alpha\beta ;\mu\nu}\;[\, G_{\alpha 3}\, ,\, G_{\beta 1}\, \pm\, i\, G_{\beta 2} \, ] \ = \ 0 \ , 
\label{eq:su4commgen3}
\EA 
from which it follows that  ${\cal B}_{\alpha\beta,2}$ and ${\cal B}_{\alpha\beta,1}$ have a piece proportional
to $\delta_{\alpha\beta}$ and a piece proportional to $d_{\alpha\beta\gamma}$. 
Rotational invariance then determines that ${\cal B}_{\alpha\beta,i}$ is a linear combination of
$\delta_{\alpha\beta}{\cal J}_i$ and  $d_{\alpha\beta\gamma}G_{\gamma i}$. The coefficients of these
terms are determined by considering the Jacobi identity of $G_{\alpha i}$ with the commutator 
in eq.~\ref{eq:su4commgen}, together with the relation among $SU(N)$ structure constants
given in Appendix~\ref{suN}. Finally, one obtains
\BA
[\, G_{\alpha i} \, ,\, G_{\beta j} \, ] \, =\, 
i\,\delta_{ij}\,f_{\alpha\beta\gamma}\, {\tilde Q}_\gamma\, +\, \frac{2}{N}\,i\,\delta_{\alpha\beta}\,\epsilon_{ijk}\,{\cal J}_k  \, + \, i\epsilon_{ijk}\,d_{\alpha\beta\gamma}\, {G}_{\gamma k} \ ,
\label{eq:su2N}
\EA
which together with 
\BA &&[\, {\tilde Q}_\alpha\, ,\, G_{\beta i}\, ]\, =\,
i\,f_{\alpha\beta\gamma} \, G_{\gamma i} \ \ \ , \ \ \
[\, {\cal J}_i \, ,\,   G_{\alpha j}  \, ] \, =\, i\,\epsilon_{ijk}\,  G_{\alpha k} \ ; \\
&&[\, {\tilde Q}_\alpha\, ,\, {\tilde Q}_\beta\, ]\, =\,
i\,f_{\alpha\beta\gamma} \, {\tilde Q}_\gamma \ \ \ , \ \ \ [\, {\cal
  J}_i \, ,\, {\cal J}_j \, ] \, =\, i\,\epsilon_{ijk}\, {\cal J}_k
\EA 
close the algebra of the symmetry group $SU(2N)$. To find the consequences
of this algebra for observable quantities like the mass-squared matrix and
the matrix elements for Goldstone boson emission and absorption, one takes matrix elements of this
algebra between hadron states $h'$ and $h$, and neglecting transitions from single-particle to multi-particle
states in the completeness sums over intermediate states, one recovers the same algebra but with the replacements
${\tilde Q}_\alpha\rightarrow \lbrack\, T_\alpha \,\rbrack_{h' h}\,$ and ${\tilde Q}_{5\alpha}\rightarrow \lbrack\, X_\alpha(\lambda) \,\rbrack_{h' h}\,$,
and corresponding replacements for $G_{\beta i}$ and ${\cal J}_k$.
This result, originally found by Weinberg~\cite{Weinberg:1994tu}, is here shown to
be a general consequence of the null-plane QCD operator algebra, valid
in any Lorentz frame.

It is important to emphasize that the $SU(2N)$ symmetry found here is
only operative in the full interacting field theory.  It is therefore
unrelated to the $SU(2N)$ invariance of the QCD Lagrangian in the
limit of no interaction.  Indeed we have show above in
section~\ref{susec:spin} that eq.~\ref{eq:OPQcomm2b}, the main
ingredient in the derivation of $SU(2N)$, in itself implies the
existence of Goldstone bosons.  In addition, in a special case, this
symmetry does emerge in a well-defined limit of QCD.  As
$\langle\,h'\,|\epsilon^\alpha|\, h\,\rangle\sim M_h-M_{h'}$, and
baryons within a given large-$N_c$ multiplet have mass splittings that
scale as $1/N_c$~\cite{Witten:1979kh}, the large-$N_c$ QCD scaling
rules suggest that for baryons $\epsilon^\alpha\sim 1/N_c$.  Of
course, as the matrix element of chiral charges between baryon states
scales as $N_c$, the $SU(2N)$ symmetry reduces to the contracted
$SU(2N)$~\cite{Weinberg:1994tu,Beane:1998xp} for baryons in the
large-$N_c$ limit, as one expects on general
grounds~\cite{Gervais:1983wq,Dashen:1993as,Dashen:1993jt}.

It is instructive to consider a simple example. Consider the case
$N=3$. Using the chiral transformation properties of the quarks,
eq.~\ref{eq:chiralassign}, one sees that a $\lambda=3/2$ baryonic
operator $\psi_{+\uparrow}\psi_{+\uparrow}\psi_{+\uparrow}$ transforms
as $({\bf 1},{\bf 1})$, $({\bf 1},{\bf 8})$, or $({\bf 1},{\bf 10})$
with respect to $SU(3)_R\otimes SU(3)_L$. Therefore, if the baryon is
a decuplet of $SU(3)_F$ with its $\lambda=3/2$ part in the $({\bf 1},{\bf 10})$, then one easily checks that its $\lambda=1/2$
part must transform as $({\bf 3},{\bf 6})$ or $({\bf 6},{\bf 3})$.
However, the different helicity states are unrelated by chiral
symmetry in itself. It is the mixed Lie-bracket,
eq.~\ref{eq:OPQcomm2b}, the expression of broken chiral symmetry in
the spin Hamiltonian, that relates the helicities. Indeed taking the
$\lambda=1/2$ decuplet to transform as $({\bf 3},{\bf 6})$ together
with an octet spin-$1/2$ field and their negative-helicity partners in
$({\bf 10},{\bf 1})\oplus({\bf 6},{\bf 3})$ together fill out the
${\bf 56}$-dimensional representation of $SU(6)$ as is familiar from
the quark model. The difference here is that this symmetry arises from
QCD symmetries and their pattern of breaking, and, in particular, has
nothing to do with the non-relativistic limit.  Hence we see that
starting from the formal null-plane QCD operator algebra, the simple
assumption that the part of the null-plane reduced Hamiltonian, $M^2$,
that breaks chiral symmetry is small implies all of the usual
consequences of the non-relativistic quark model, without the need of
any further assumption like the existence of constituent quark degrees
of freedom~\cite{Weinberg:1994tu}.

%
\section{Conclusion}
\label{sec:conc}

\noindent Usually one views the spontaneous breaking of a symmetry as
the non-invariance of the vacuum state with respect to the
symmetry. However, in relativistic theories of quantum mechanics, this
picture is purely a matter of convention. We have seen that the
front-form vacuum is a singlet with respect to all symmetries and yet
spontaneous symmetry breaking can occur via non-conserved currents
whose divergences are directly proportional to S-matrix elements for
the emission and absorption of Goldstone bosons.  One may view the
null-plane description as a change of coordinates which moves
dynamical information out of the vacuum state and into the interaction
operators of the theory. The primary advantage of working with the
null-plane description is that broken chiral symmetry constraints
become manifest in the sense that there are non-trivial Lie brackets
between the Poincar\'e generators and the broken symmetry
generators. In the instant-form, the chiral constraints that appear
naturally in the front-form are present, but require one to work in
special Lorentz frames and to make assumptions about the asymptotic
behavior of Goldstone-boson scattering amplitudes.
\vskip0.2in

Here we will restate the main conclusions of this paper:

\vskip0.2in
\noindent$\bullet$ In the front-form, spontaneous chiral symmetry
breaking is contained entirely in the three null-plane reduced
Hamiltonians, which encode the mass spectrum and spin content of a
given theory. This must be the case as the null-plane chiral charges
annihilate the vacuum state, and therefore chiral symmetry breaking
cannot be attributed to the formation of chiral-symmetry breaking
condensates. In null-plane QCD, all chiral symmetry breaking
arises from the symmetry breaking parts of the reduced Hamiltonians,
given explicitly in eqs.~\ref{eq:symmbreakhamsa} and \ref{eq:symmbreakhamsb}.

\vskip0.2in
\noindent$\bullet$ Goldstone's theorem on the null-plane follows
directly from the Lie-brackets between the null-plane Hamiltonians and
the chiral charges. A consistent null-plane description of spontaneous
symmetry breaking requires that a small explicit symmetry-breaking
operator be included and that this explicit symmetry breaking be taken
to zero only at the level of matrix elements of operators. The
divergence of the axial-vector current is proportional to the explicit
symmetry breaking. Therefore, as the current cannot be conserved in
the symmetry limit, the existence of massless states arises as a
consequence of the need to cancel the explicit breaking parameter that
appears in its divergence.

\vskip0.2in
\noindent$\bullet$ The Gell-Mann-Oakes-Renner relation is recovered in
null-plane QCD and a general prescription exists for translating all
chiral-symmetry breaking condensates in instant-form QCD to
chiral-singlet condensates in null-plane QCD.  It is therefore
simplistic to say that the vacuum is trivial in the front-form, since
there are necessarily symmetry-preserving condensates which arise from
modes with strictly zero longitudinal momentum. In particular,
in contrast with claims in the literature~\cite{Brodsky:2008xm,Brodsky:2008xu,Brodsky:2009zd,Brodsky:2010xf},
we expect that the QCD vacuum energy is unaltered in moving from
the instant-form to the front-form descriptions of QCD, as
is essential for the consistency of null-plane QCD.

\vskip0.2in
\noindent$\bullet$ A simple solution of the null-plane operator
algebra recovers the spin-flavor symmetry of the constituent quark
model.  This result was obtained originally in
Ref.~\cite{Weinberg:1994tu}, which obtained the algebra of charges and
Hamiltonians by working with sum rules obtained in special Lorentz
frames, and using input from Regge-pole theory expectations of the
asymptotic behavior of scattering amplitudes involving Goldstone
bosons.  The results of the present work may be viewed as an attempt to clarify this
original work by formulating it in a Lorentz frame-independent manner
which follows directly from null-plane QCD.

\vskip0.2in

In the null-plane formulation of QCD, the loss of manifest Lorentz
invariance and locality are, operationally, a result of integrating
out non-dynamical degrees of freedom. Physically, it is clear that the
loss of Lorentz invariance is tied to the fact that the essence of
Lorentz invariance lies in the Poincar\'e Lie brackets that must be
satisfied by the spin generators, and, of course, on the null-plane
spin is dynamical and therefore requires the solution of the theory to
properly implement. By contrast, the non-locality of the theory would
appear to be related to the fact that the null-plane chiral symmetry
constraints on observables are properly formulated as sum rules which
span many energy scales, and therefore do not exhibit the separation
of scales that allows a useful description in terms of local
Lagrangian effective field theory.  Indeed, it appears that, in some
sense, scattering amplitudes are the fundamental objects in the
null-plane formulation. This is particularly clear from the
Lie-brackets that mix the Poincar\'e and chiral symmetry generators,
which are given by the S-matrix elements for Goldstone boson emission
and absorption.  From a theoretical standpoint, the most interesting
consequences of the results obtained in this paper are apparent only
in the large-$N_c$ limit, which will be treated separately.

\vskip0.3in

\noindent 
I thank Ulf-G.~Mei\ss ner for valuable comments on the manuscript, and
T.~Becher, G.~Colangelo, H.~Leutwyler, F.~Niedermayer, and U.~Wenger
for useful discussions. I am particularly grateful to the Institute
for Theoretical Physics at the University of Bern for providing a
stimulating work environment during academic year 2010/2011.  The
Albert Einstein Center for Fundamental Physics is supported by the
“Innovations- und Kooperationsprojekt C-13” of the “Schweizerische
Universit\"atskonferenz SUK/CRUS”. I gratefully acknowledge the
hospitality of HISKP-theorie and the support of the Mercator programme
of the Deutsche Forschungsgemeinschaft during academic year
2012/2013. This work was supported in part by NSF CAREER Grant
PHY-0645570 and continuing grant PHY1206498.


%

\appendix

\section{Null-plane conventions}\label{npconventions}

\noindent We adopt the metric convention:
\BA
       g^{\mu\nu}\ =\ g_{\mu\nu}\ =\ \begin{pmatrix}1&0&0&0 \cr 
                                      0&-1&0&0 \cr 
                                      0&0&-1&0 \cr 
                                      0&0&0&-1 \cr \end{pmatrix} 
\label{eq:ETmetric}
\EA which takes the contravariant coordinate four-vector
$x^\mu=(x^0,x^1,x^2,x^3)=(t,x,y,z)$ to the covariant coordinate
four-vector $x_\mu=g_{\mu\nu}x^\mu=(x_0,-x_1,-x_2,-x_3)$. With
$x^+ \equiv x \cdot n$ and $x^- \equiv x \cdot n^\ast$, we
denote the null-plane contravariant coordinate four-vector
by ${\tilde x}^\mu=(x^+,x^1,x^2,x^-)$. Then we have
\BA
{\tilde x}^\mu\ = \ {\cal C}^\mu_\nu\ x^\mu \ ,
\label{eq:ETtoLCapp}
\EA
with
\BA
{\cal C}^\mu_\nu\ =\         \begin{pmatrix} {1}/\textstyle{\sqrt{2}}&0&0&1/\textstyle{\sqrt{2}} \cr 
                                      0&1&0&0 \cr 
                                      0&0&1&0 \cr 
                                      1/\textstyle{\sqrt{2}}&0&0&-1/\textstyle{\sqrt{2}} \cr \end{pmatrix} \ .
\label{eq:ETtoLC2}
\EA
This matrix transforms all Lorentz tensors in the
instant-form notation to the front-form notation. For instance,
the null-plane metric tensor is given by
\BA
{\tilde g}_{\mu\nu}\ =\ ({\cal C}^{-1})^\alpha_\mu\ g_{\alpha\beta}\ ({\cal C}^{-1})^\beta_\nu
\label{eq:LCmetric1app}
\EA
which gives 
\BA
       {\tilde g}^{\mu\nu}\ =\ {\tilde g}_{\mu\nu}\ =\ \begin{pmatrix}0&0&0&1 \cr 
                                                            0&-1&0&0 \cr 
                                                            0&0&-1&0 \cr 
                                                            1&0&0&0 \cr \end{pmatrix} \ .
\label{eq:LCmetric2}
\EA 
We can now form the scalar product
\BA
       x\cdot p \, =\, x^\mu p_\mu \, =\, x^+ p_+ + x^- p_- + x^1 p_1+ x^2 p_2 \, =\,  x^+ p^- + x^- p^+
       - {\bf x}_\perp\cdot{\bf p}_\perp \ .
\label{eq:LCsp}
\EA 
The indices $i,j,k,\dots$ are spatial indices that range over
$1,2,3$, and $r,s,t,\ldots$ are transverse indices that range over
$1,2$. We place all transverse coordinates, momenta and fields in boldface,
and additionally label coordinates and momenta with the $\perp$ symbol.
The totally antisymmetric symbol is
\BA
   \epsilon_{+12}^{\phantom{+12}+}\, =\, 1\, =\,
   \epsilon_{+12-}=1 \ .
\label{eq:LCas}
\EA
Note that $ \partial _+ = \partial ^- $ is a time-like derivative
$\partial /\partial x ^+ = \partial /\partial x _-$ as opposed to $
\partial _- = \partial ^+ $, which is a space-like derivative
$\partial /\partial x ^- = \partial /\partial x _+$. Many more
useful relations can be found in Ref.~\cite{Brodsky:1997de}.

\section{Free fermion fields decomposed}\label{FFdecompose}

\noindent Consider the Lagrangian of a free fermion of mass $m$,
\begin{equation}
{\cal L}(x)
=
 \bar\psi (x)\Big\lbrack\ft{i}2\left(
\stackrel{{}_\rightarrow}{\partial}_\mu - \stackrel{{}_\leftarrow}{\partial}_\mu
\right)
\gamma^\mu  - m \Big\rbrack\psi (x) \ .
\label{eq:freefermion}
\end{equation}
The Dirac equations of motion for the fermion and anti-fermion fields are:
\begin{equation}
\left( i \gamma^\mu\stackrel{{}_\rightarrow}{\partial}_\mu - m\right)\psi(x)\ =\ 0
\qquad , \qquad
 \bar\psi (x)\left(i \gamma^\mu\stackrel{{}_\leftarrow}{\partial}_\mu + m\right)\ =\ 0 \ .
\label{eq:freefermiondirac}
\end{equation}
In order to express the Lagrangian in null-plane coordinates such that the null-plane dispersion relation is recovered, 
the fermion field is decomposed into two components,
\BA
\psi = {\mit\Pi}^+ \psi + {\mit\Pi}^- \psi
\equiv
\psi_+ + \psi_-
\, ,
\label{eq:decap}
\EA
where the projection operator is defined as ${\mit\Pi}^{\pm} = \ft12 \gamma^{\mp} \gamma^{\pm}$, with
\BA
\gamma^+ \equiv \gamma \cdot n = \ft{1}{\sqrt{2}} (\gamma^0 + \gamma^3)
\ \  , \ \
\gamma^- \equiv \gamma \cdot n^\ast = \ft{1}{\sqrt{2}} (\gamma^0 - \gamma^3) \ .
\label{eq:NPgammas}
\EA
Application of the projection operator to the Dirac equation then gives
\begin{equation}
{2 i\! \stackrel{{}_\rightarrow}{\partial^+}}\psi_- \ =\ \left( -i \bit{\gamma}^r\!\stackrel{{}_\rightarrow}{{\partial}^r}\! \ +\ m \right) \gamma^+\psi_+
\quad , \quad
2 i\, \psi^\dagger_- {\stackrel{{}_\leftarrow}{\partial^+}}\ =\  \psi^\dagger_+ \gamma^- \left( i {\gamma}^r\!\stackrel{{}_\leftarrow}{{\partial}^r}\! \ -\ m \right) \ ,
\end{equation}
which reveals that the $\psi_-$ field is non-dynamical. One can solve for $\psi_-$ by inverting the longitudinal coordinate derivative operator to give
\begin{equation}
\psi_- \ =\ \frac{1}{2 i\! \stackrel{{}_\rightarrow}{\partial^+}}\left( -i \bit{\gamma}^r\!\stackrel{{}_\rightarrow}{{\partial}^r}\! \ +\ m \right) \gamma^+\psi_+
\quad , \quad
\psi^\dagger_- \ =\  \psi^\dagger_+ \gamma^- \left( i \bit{\gamma}^r\!\stackrel{{}_\leftarrow}{{\partial}^r}\! \ -\ m \right) \frac{1}{2 i\! \stackrel{{}_\leftarrow}{\partial^+}} \ ,
\label{eq:constrdirac}
\end{equation}
where $(1/\partial^+)\partial^+=\partial^+(1/\partial^+)=1$. An explicit representation of this operator can be taken as:
\begin{equation}
\left(\,\frac{1}{\partial^+}\,\right)\, f\left(x^+, x^- , {\bf x}_\perp\right)\ =\ \frac{1}{4}\,\int_{-\infty}^{+\infty}\, dy^-\, \epsilon\left(x^-\ -\ y^-\right)\, 
f\left(x^+, y^- , {\bf x}_\perp\right) \ ,
\end{equation}
where $\epsilon(z)=-1,0,1$ for $x>0,=0,<0$, respectively. Now, using eq.~\ref{eq:decap} and the constraint equation, eq.~\ref{eq:constrdirac},
gives the null-plane free-fermion Lagrangian,
\begin{equation}
{\tilde{\cal L}}(x)
=-\psi_+^\dagger(x)\, \frac{\Box^2+m^2}{\sqrt{2}i\partial^+}\,\psi_+(x) \ ,
 \label{eq:freefermionNP}
\end{equation}
where $\Box^2 \equiv 2\partial^+\partial^--\partial^r\partial^r$.

It is useful to list the Poincar\'e generators in the free fermion theory. 
We take
\begin{equation}
{T}^{\mu\nu}\ =\ - g^{\mu\nu} {\cal L}
\ +\ 
\frac{i}{2}
\bar\psi\gamma^\nu
\stackrel{{}_\leftrightarrow}{\partial}{\!}^\mu 
\psi
\, 
\label{EMtensorfree}
\end{equation}
as the free-fermion energy-momentum tensor.
The free-fermion Poincar\'e generators are then obtained via
\begin{eqnarray}
{\tilde P}^\mu & =& \int\, d x^-\, d^2 \bit{x}_\perp {T}^{\mu +} \ ; \\
{\tilde M}^{\mu\nu} & =& \int\, d x^-\, d^2 \bit{x}_\perp \left( x^\mu {T}^{\nu +}- x^\nu {T}^{\mu +}+
\textstyle{\frac{1}{4}}\bar\psi\lbrace \gamma^+,\sigma^{\mu\nu}\rbrace \psi\right) \ ,
\label{freepoincarePandM}
\end{eqnarray}
where $\sigma^{\mu\nu}=i[\gamma^\mu,\gamma^\nu ]/2$. The free-fermion stability group generators are~\cite{Eichten:1973ip}:
\begin{eqnarray}
P^r &=& i\sqrt{2}\;\int\, d x^-\, d^2 \bit{x}_\perp\, \psi_+^\dagger(x)\,\partial^r\,\psi_+(x) \ ; \label{eq:kinfermp}\\
P^+ &=& i\sqrt{2}\;\int\, d x^-\, d^2 \bit{x}_\perp\, \psi_+^\dagger(x)\,\partial^+\,\psi_+(x) \ ; \label{eq:kinfermpp}\\
E^r &=& i\sqrt{2}\;\int\, d x^-\, d^2 \bit{x}_\perp\, \psi_+^\dagger(x)\,\left(x^r\partial^+-x^+\partial^r\right)\,\psi_+(x) \ ; \label{eq:kinferme}\\
K^3 &=& i\sqrt{2}\;\int\, d x^-\, d^2 \bit{x}_\perp\, \psi_+^\dagger(x)\,\Big\lbrack -x^+\frac{1}{2\partial^+}\left(-\partial^r\partial^r+m^2\right)
-x^-\partial^+-\frac{1}{2} \Big\rbrack\,\psi_+(x) ; \label{eq:kinfermk}\\
J^3 &=& i\sqrt{2}\;\int\, d x^-\, d^2 \bit{x}_\perp\, \psi_+^\dagger(x)\,\epsilon^{rs}\left(x^r\partial^s+\textstyle{\frac{1}{4}}\gamma^r\gamma^s\right)\,\psi_+(x) \ ,
\label{eq:kinfermj}
\end{eqnarray}
and the Hamiltonians are:
\begin{eqnarray}
P^- &=& i\sqrt{2}\;\int\, d x^-\, d^2 \bit{x}_\perp\, \psi_+^\dagger(x)\frac{1}{2\partial^+}\left(-\partial^r\partial^r+m^2\right)\,\psi_+(x) \ ; 
\label{freenergy} \\
F^r &=& i\sqrt{2}\;\int\, d x^-\, d^2 \bit{x}_\perp\, \psi_+^\dagger(x)\,\Big\lbrack -x^r\frac{1}{2\partial^+}\left(-\partial^r\partial^r+m^2\right) -x^-\partial^r \\
&&\hskip18.3em-\frac{\gamma^r}{2\partial^+}\left(-\gamma^s\partial^s+im \right) \Big\rbrack\,\psi_+(x) .
\label{freespin}
\end{eqnarray}
It is clear that the null-plane dispersion relation, eq.~\ref{eq:LCdisp}, is correctly reproduced by eq.~\ref{freenergy}.

The dynamical fermion field $\psi_+$ can be expressed in momentum space as
\begin{eqnarray}
\hskip-1.3em \psi_+ (x) = \sum_{\lambda = \uparrow\downarrow}
\int \frac{d k^+ d^2 \bit{k}_\perp}{2 k^+ (2 \pi)^3}
 \left\{ b_\lambda (k^+ , \bit{k}_\perp) u_+(k,\lambda) {\rm e}^{- i k\cdot x} +
d^\dagger_\lambda (k^+ , \bit{k}_\perp) v_+(k,\lambda) {\rm e}^{i k\cdot x} \right\} \ ,
\label{eq:freediracmomspace}
\end{eqnarray}
where $b_\lambda (k^+ , \bit{k}_\perp)$ destroys a fermion and
$d^\dagger_\lambda (k^+ , \bit{k}_\perp)$ creates an antifermion. This
decomposition is meaningful only on the initial surface, $x^+=0$,
where the fermions are free. The creation/destruction operators satisfy
the anti-commutation relations
\begin{eqnarray}
\{
b_\lambda (k^+ , \bit{k}_\perp) ,
b^\dagger_{\lambda'} (k^{\prime +}, \bit{k}^\prime_\perp )
\}
&=&
2 k^+ (2 \pi)^3
\delta (k^+ - k^{\prime +})
\delta^{2} (\bit{k}_\perp - \bit{k}'_\perp)
\delta_{\lambda \lambda^\prime}
\, ; \\
{}[ \,
d_\lambda (k^+ , \bit{k}_\perp) ,
d^\dagger_{\lambda'} (k^{\prime +}, \bit{k}^\prime_\perp ) \,
{}]
&=&
2 k^+ (2 \pi)^3
\delta (k^+ - k^{\prime +})
\delta^{2} (\bit{k}_\perp - \bit{k}'_\perp)
\delta_{\lambda \lambda^\prime}
\, .
\end{eqnarray}
which in turn imply that the fermion field $\psi_+$ satisfies
\begin{eqnarray}
\{ \psi_+(x)\, ,\, \psi^\dagger_+(y)\}|_{x^+=y^+}
\ = \
\ft{1}{\sqrt{2}} {\mit\Pi}^+\;
\delta (x^- - y^-)
\delta^{2} (\bit{x}_\perp - \bit{y}_\perp) \ .
\end{eqnarray}

The solutions of the free Dirac equation in the chiral representation of the gamma matrices 
are~\cite{Kogut:1969xa}:

\BA
u(k,\uparrow) \ =\ \frac{1}{2^{1/4}\sqrt{k^+}}\begin{pmatrix}&\sqrt{2}k^+ \cr 
                                                            &k_\perp \cr 
                                                            &m \cr 
                                                            &0 \end{pmatrix} \ \ \ \ , \ \ \ \
u(k,\downarrow) \ =\ \frac{1}{2^{1/4}\sqrt{k^+}}\begin{pmatrix}&0 \cr 
                                                            &m \cr 
                                                            &-\bar{k}_\perp \cr 
                                                            &\sqrt{2}k^+ \end{pmatrix} \ ; \\
v(k,\uparrow) \ =\ \frac{1}{2^{1/4}\sqrt{k^+}}\begin{pmatrix}&0 \cr 
                                                            &-m \cr 
                                                            &-\bar{k}_\perp \cr 
                                                            &\sqrt{2}k^+ \end{pmatrix} \ \ \ \ , \ \ \ \
v(k,\downarrow) \ =\ \frac{1}{2^{1/4}\sqrt{k^+}}\begin{pmatrix}&\sqrt{2}k^+ \cr 
                                                            &k_\perp \cr 
                                                            &-m \cr 
                                                            &0 \end{pmatrix} \ ,
\label{eq:uandvdef}
\EA 
where $k_\perp\equiv k_1+ik_2$ and $\bar{k}_\perp\equiv k_1-ik_2$. Projecting out
the dynamical spinors gives
\BA
&u_+(k,\uparrow) \ =\ \Pi^+ u(k,\uparrow) \ =\ {2^{1/4}\sqrt{k^+}}\begin{pmatrix}&1 \cr 
                                                            &0 \cr 
                                                            &0 \cr 
                                                            &0 \end{pmatrix} \ = \ v_+(k,\downarrow)\ =\ \Pi^+ v(k,\downarrow) \ ;\\
&u_+(k,\downarrow) \ =\ \Pi^+ u(k,\downarrow) \ =\ {2^{1/4}\sqrt{k^+}}\begin{pmatrix}&0 \cr 
                                                            &0 \cr 
                                                            &0 \cr 
                                                            &1 \end{pmatrix} \ = \ v_+(k,\uparrow)\ =\ \Pi^+ v(k,\uparrow) \ ,
\label{eq:uandvdefeigen}
\EA 
which leads to the eigenvalue equations,
\BA
u^\dagger_+(k,\lambda) \gamma_5 u_+(k,\lambda) \ &=&\ u^\dagger_+(k,\lambda) 2\Sigma_{12} u_+(k,\lambda) \ =\ 2\lambda\sqrt{2} k^+ \ ;\\
v^\dagger_+(k,\lambda) \gamma_5 v_+(k,\lambda) \ &=&\ v^\dagger_+(k,\lambda) 2\Sigma_{12} v_+(k,\lambda)\ =\ -2\lambda\sqrt{2} k^+ \ ,
\label{eq:heleigen}
\EA 
where $\Sigma_{12}\equiv\gamma_1 \gamma_2/2$.  The relation between
chirality and helicity in the null-plane formulation arises from these
relations which arise from the fact that each of the fields has only
a single non-vanishing component.  Now it is a straightforward matter
to express the Poincar\'e generators in the momentum-space
representation. For instance, comparing eqs.~\ref{eq:J3eigen},
\ref{eq:J3atrest}, and \ref{eq:kinfermj} gives the free-fermion
helicity operator,
\begin{eqnarray}
{\cal J}^3 &=& i\sqrt{2}\;\int\, d x^-\, d^2 \bit{x}_\perp\, \psi_+^\dagger(x)\,\Sigma^{12}\,\psi_+(x) \ ,
\end{eqnarray}
which, using eqs.~\ref{eq:freediracmomspace} and ~\ref{eq:heleigen}, is found to have the momentum-space representation
\begin{eqnarray}
{\cal J}^3  &=& \sum_{\lambda = \uparrow\downarrow}\lambda
\int \frac{d k^+ d^2 \bit{k}_\perp}{2 k^+ (2 \pi)^3}
 \left\{  b^\dagger_\lambda (k^+ , \bit{k}_\perp) b_\lambda (k^+ , \bit{k}_\perp)   +
d^\dagger_\lambda (k^+ , \bit{k}_\perp) d_\lambda (k^+ , \bit{k}_\perp) \right\} \ .
\label{eq:freehelquark}
\end{eqnarray}
This operator explicitly counts the helicity of the fermions and the antifermions.

\section{Free gauge fields decomposed}\label{FGdecompose}

\noindent Consider the Lagrangian of a free gluon field,
\BA
{\cal L}(x) \ =\
- \ft14 F_{\mu\nu}^a (x) F^{\mu\nu}_a (x) \ .
\label{eq:freegauge}
\EA
The equation of motion is
\BA
D_\mu^{ab}\;F^{\mu\nu}_b\ =\ 0 \ ,
\label{eq:freegaugeEOM}
\EA
where $D_\mu^{ab}=\delta^{ab}\partial_\mu + g\;f^{acb}A^c_\mu$, 
where here $f^{acb}$ are $SU(3)$ structure constants. 
The gauge potential can be expressed in null-plane coordinates as
\begin{equation}
A^\mu = (A^+,\bit{A},A^-)
\end{equation}
where $A^+=n\cdot A$, $A^-=n^*\cdot A$ and $\bit{A} = (A^1, A^2)$.
Working in light-cone gauge, $A^+=0$, one finds
\BA
\partial^+ A^-_a \ =\ -\frac{1}{\partial^+}\; D_r^{ab}\partial^+ \bit{A}^r_b \ .
\label{eq:freegaugeEOMsol1}
\EA
Therefore $A^-_a$ is non-dynamical and can be integrated out, giving
\BA
{\cal L}(x) \ =\ 
-
\ft{1}{4} \bit{F}_{a}^{rs} \bit{F}_{a}^{rs}
\ +\ 
\left( \partial^+ \bit{A}^r_{a} \right)
\left( \partial^- \bit{A}^r_{a} \right)
-\ft12
\left(
\frac{1}{\partial^+}
D^r_{ab} \partial^+ \bit{A}^r_{b}\right)^2
\ .
\label{eq:freegaugenl}
\EA
The light-cone gauge does not fix the gauge entirely and therefore
to eliminate all redundancy one should assign a boundary condition
to the transverse gauge field; e.g. $\bit{A}^r_{a}(x^+,{\bf x}_\perp,x^-=\infty)=0$.

\section{$SU(N)$  conventions}\label{suN}

\noindent The fundamental representation $SU(N)$ generators $T_\alpha$ with $\alpha=1,\ldots,N^2-1$ satisfy:
\begin{eqnarray}
[\, T_\alpha \, ,\,  T_\beta \, ]\, &=& \, i\,f_{\alpha\beta\gamma} \ ; \\
\{\, T_\alpha \, ,\,  T_\beta \, \}\, &=& \, \frac{1}{N}{\bf 1}\,\delta_{\alpha\beta} \ +\ d_{\alpha\beta\gamma}\,T_\gamma \ ,
\end{eqnarray}
where ${\bf 1}$ is the $N\times N$ unit matrix, and hence are normalized such that ${\rm Tr}(T_\alpha T_\beta)=\delta_{\alpha\beta}/2$.
The structure constants satisfy the relations:
\begin{eqnarray}
f_{\alpha\mu\nu}f_{\beta\mu\nu} \ &=&\ N\,\delta_{\alpha\beta} \ ; \\
d_{\alpha\mu\nu}d_{\beta\mu\nu} \ &=&\ \frac{N^2\; -\;4}{N}\,\delta_{\alpha\beta} \ . 
\end{eqnarray}
An additional useful relation is:
\begin{eqnarray}
f_{\alpha\beta\nu}f_{\gamma\mu\nu} \ &=&\ \frac{2}{N}\,\left(\,\delta_{\alpha\gamma}\,\delta_{\beta\mu}\,-\, \delta_{\alpha\mu}\,\delta_{\beta\gamma}\,\right)\ +\ 
d_{\alpha\gamma\nu}d_{\beta\mu\nu} \ -\ d_{\beta\gamma\nu}d_{\alpha\mu\nu} \ .
\end{eqnarray}

\vfill\eject
\bibliographystyle{JHEP}
\bibliography{bibi}

\providecommand{\href}[2]{#2}\begingroup\raggedright\begin{thebibliography}{10}

\bibitem{Dirac:1949cp}
P.~A. Dirac, {\it {Forms of Relativistic Dynamics}},  {\em Rev.Mod.Phys.} {\bf
  21} (1949) 392--399.

\bibitem{Weinberg:1966jm}
S.~Weinberg, {\it {Dynamics at infinite momentum}},  {\em Phys.Rev.} {\bf 150}
  (1966) 1313--1318.

\bibitem{Susskind:1967rg}
L.~Susskind, {\it {Model of selfinduced strong interactions}},  {\em Phys.Rev.}
  {\bf 165} (1968) 1535--1546.

\bibitem{Bardakci:1969dv}
K.~Bardakci and M.~Halpern, {\it {Theories at infinite momentum}},  {\em
  Phys.Rev.} {\bf 176} (1968) 1686--1699.

\bibitem{Kogut:1969xa}
J.~B. Kogut and D.~E. Soper, {\it {Quantum Electrodynamics in the Infinite
  Momentum Frame}},  {\em Phys.Rev.} {\bf D1} (1970) 2901--2913.

\bibitem{Leutwyler:1977vy}
H.~Leutwyler and J.~Stern, {\it {Relativistic Dynamics on a Null Plane}},  {\em
  Annals Phys.} {\bf 112} (1978) 94.

\bibitem{Weinberg:1969hw}
S.~Weinberg, {\it {Algebraic realizations of chiral symmetry}},  {\em
  Phys.Rev.} {\bf 177} (1969) 2604--2620.

\bibitem{Weinberg:1969db}
S.~Weinberg, {\it {Algebraic structure of superconvergence relations}},  {\em
  Phys.Rev.Lett.} {\bf 22} (1969) 1023--1025.

\bibitem{Weinberg:1990xn}
S.~Weinberg, {\it {Mended symmetries}},  {\em Phys.Rev.Lett.} {\bf 65} (1990)
  1177--1180.

\bibitem{Weinberg:1994tu}
S.~Weinberg, {\it {Strong interactions at low-energies}},
  \href{http://xxx.lanl.gov/abs/hep-ph/9412326}{{\tt hep-ph/9412326}}.

\bibitem{Jersak:1969zg}
J.~Jersak and J.~Stern, {\it {Algebra of currents on the light cone}},  {\em
  Nuovo Cim.} {\bf A59} (1969) 315--327.

\bibitem{Leutwyler:1969av}
H.~Leutwyler, {\it {Current algebra and lightlike charges}},  {\em Springer
  Tracts Mod.Phys.} {\bf 50} (1969) 29--41.

\bibitem{Feinberg:1973qb}
F.~Feinberg, {\it {Algebraic realization of chiral symmetry on the light
  cone}},  {\em Phys.Rev.} {\bf D7} (1973) 540--549.

\bibitem{Eichten:1973ip}
E.~Eichten, F.~Feinberg, and J.~Willemsen, {\it {Current and consituent quarks
  in the light-cone quantization}},  {\em Phys.Rev.} {\bf D8} (1973)
  1204--1219.

\bibitem{Casher:1973vh}
A.~Casher and L.~Susskind, {\it {A quark model of mesons based on chiral
  symmetry}},  {\em Phys.Lett.} {\bf B44} (1973) 171--174.

\bibitem{Casher:1974xd}
A.~Casher and L.~Susskind, {\it {Chiral magnetism (or magnetohadrochironics)}},
   {\em Phys.Rev.} {\bf D9} (1974) 436--460.

\bibitem{Carlitz:1974sg}
R.~D. Carlitz, D.~Heckathorn, J.~Kaur, and W.-K. Tung, {\it {Chiral Symmetry
  and the Quark Model}},  {\em Phys.Rev.} {\bf D11} (1975) 1234.

\bibitem{Sazdjian:1974gk}
H.~Sazdjian and J.~Stern, {\it {Traces of Chiral Symmetry on Light Planes}},
  {\em Nucl.Phys.} {\bf B94} (1975) 163.

\bibitem{Wilson:1994fk}
K.~G. Wilson, T.~S. Walhout, A.~Harindranath, W.-M. Zhang, R.~J. Perry, et~al.,
  {\it {Nonperturbative QCD: A Weak coupling treatment on the light front}},
  {\em Phys.Rev.} {\bf D49} (1994) 6720--6766,
  [\href{http://xxx.lanl.gov/abs/hep-th/9401153}{{\tt hep-th/9401153}}].

\bibitem{Wilson:1994gn}
K.~G. Wilson and M.~M. Brisudova, {\it {Chiral symmetry breaking and light
  front QCD}},  \href{http://xxx.lanl.gov/abs/hep-th/9411008}{{\tt
  hep-th/9411008}}.

\bibitem{Susskind:1994wr}
L.~Susskind and M.~Burkardt, {\it {A Model of mesons based on chi(SB) in the
  light front frame}},  \href{http://xxx.lanl.gov/abs/hep-ph/9410313}{{\tt
  hep-ph/9410313}}.

\bibitem{Kim:1994rm}
Y.~Kim, S.~Tsujimaru, and K.~Yamawaki, {\it {Can the Nambu-Goldstone boson live
  on the light front?}},  {\em Phys.Rev.Lett.} {\bf 74} (1995) 4771--4774,
  [\href{http://xxx.lanl.gov/abs/hep-ph/9502214}{{\tt hep-ph/9502214}}].

\bibitem{Burkardt:1995eb}
M.~Burkardt, {\it {Trivial vacua, high orders in perturbation theory and
  nontrivial condensates}},  {\em Phys.Rev.} {\bf D53} (1996) 933--938,
  [\href{http://xxx.lanl.gov/abs/hep-ph/9509226}{{\tt hep-ph/9509226}}].

\bibitem{Burkardt:1996pa}
M.~Burkardt and H.~El-Khozondar, {\it {A (3+1)-dimensional light front model
  with spontaneous breaking of chiral symmetry}},  {\em Phys.Rev.} {\bf D55}
  (1997) 6514--6521, [\href{http://xxx.lanl.gov/abs/hep-ph/9609250}{{\tt
  hep-ph/9609250}}].

\bibitem{Burkardt:1998dd}
M.~Burkardt, {\it {Dynamical vertex mass generation and chiral symmetry
  breaking on the light front}},  {\em Phys.Rev.} {\bf D58} (1998) 096015,
  [\href{http://xxx.lanl.gov/abs/hep-th/9805088}{{\tt hep-th/9805088}}].

\bibitem{Yamawaki:1998cy}
K.~Yamawaki, {\it {Zero mode problem on the light front}},
  \href{http://xxx.lanl.gov/abs/hep-th/9802037}{{\tt hep-th/9802037}}.

\bibitem{Itakura:2001yt}
K.~Itakura and S.~Maedan, {\it {Light front realization of chiral symmetry
  breaking}},  {\em Prog.Theor.Phys.} {\bf 105} (2001) 537--571,
  [\href{http://xxx.lanl.gov/abs/hep-ph/0102330}{{\tt hep-ph/0102330}}].

\bibitem{Burkardt:2002yf}
M.~Burkardt, F.~Lenz, and M.~Thies, {\it {Chiral condensate and short time
  evolution of QCD(1+1) on the light cone}},  {\em Phys.Rev.} {\bf D65} (2002)
  125002, [\href{http://xxx.lanl.gov/abs/hep-th/0201235}{{\tt
  hep-th/0201235}}].

\bibitem{Wu:2003vn}
M.-H. Wu and W.-M. Zhang, {\it {Chiral symmetry in light front QCD}},  {\em
  JHEP} {\bf 0404} (2004) 045,
  [\href{http://xxx.lanl.gov/abs/hep-ph/0310095}{{\tt hep-ph/0310095}}].

\bibitem{Lenz:2004tw}
F.~Lenz, K.~Ohta, M.~Thies, and K.~Yazaki, {\it {Chiral symmetry in light cone
  field theory}},  {\em Phys.Rev.} {\bf D70} (2004) 025015,
  [\href{http://xxx.lanl.gov/abs/hep-th/0403186}{{\tt hep-th/0403186}}].

\bibitem{Dalley:2004re}
S.~Dalley and G.~McCartor, {\it {Spontaneously broken quark helicity
  symmetry}},  {\em Annals Phys.} {\bf 321} (2006) 402--420,
  [\href{http://xxx.lanl.gov/abs/hep-ph/0406287}{{\tt hep-ph/0406287}}].

\bibitem{Brodsky:2008xm}
S.~J. Brodsky and R.~Shrock, {\it {On Condensates in Strongly Coupled Gauge
  Theories}},  {\em Proc.Nat.Acad.Sci.} {\bf 108} (2011) 45--50,
  [\href{http://xxx.lanl.gov/abs/0803.2541}{{\tt arXiv:0803.2541}}].

\bibitem{Brodsky:2008xu}
S.~J. Brodsky and R.~Shrock, {\it {Standard-Model Condensates and the
  Cosmological Constant}},  {\em Proc.Nat.Acad.Sci.} {\bf 108} (2011) 45--50,
  [\href{http://xxx.lanl.gov/abs/0803.2554}{{\tt arXiv:0803.2554}}].

\bibitem{Brodsky:2009zd}
S.~J. Brodsky and R.~Shrock, {\it {Condensates in Quantum Chromodynamics and
  the Cosmological Constant}},  {\em Proc.Nat.Acad.Sci.} {\bf 108} (2011)
  45--50, [\href{http://xxx.lanl.gov/abs/0905.1151}{{\tt arXiv:0905.1151}}].

\bibitem{Ji:2009jc}
C.-R. Ji, W.~Melnitchouk, and A.~Thomas, {\it {Equivalence of pion loops in
  equal-time and light-front dynamics}},  {\em Phys.Rev.} {\bf D80} (2009)
  054018, [\href{http://xxx.lanl.gov/abs/0906.3497}{{\tt arXiv:0906.3497}}].

\bibitem{Strikman:2010pu}
M.~Strikman and C.~Weiss, {\it {Quantifying the nucleon's pion cloud with
  transverse charge densities}},  {\em Phys.Rev.} {\bf C82} (2010) 042201,
  [\href{http://xxx.lanl.gov/abs/1004.3535}{{\tt arXiv:1004.3535}}].

\bibitem{Chang:2011mu}
L.~Chang, C.~D. Roberts, and P.~C. Tandy, {\it {Expanding the concept of
  in-hadron condensates}},  {\em Phys.Rev.} {\bf C85} (2012) 012201,
  [\href{http://xxx.lanl.gov/abs/1109.2903}{{\tt arXiv:1109.2903}}].

\bibitem{Alberg:2012wr}
M.~Alberg and G.~A. Miller, {\it {Taming the Pion Cloud of the Nucleon}},  {\em
  Phys.Rev.Lett.} {\bf 108} (2012) 172001,
  [\href{http://xxx.lanl.gov/abs/1201.4184}{{\tt arXiv:1201.4184}}].

\bibitem{Brodsky:2010xf}
S.~J. Brodsky, C.~D. Roberts, R.~Shrock, and P.~C. Tandy, {\it {Essence of the
  vacuum quark condensate}},  {\em Phys.Rev.} {\bf C82} (2010) 022201,
  [\href{http://xxx.lanl.gov/abs/1005.4610}{{\tt arXiv:1005.4610}}].

\bibitem{Schweitzer:2012hh}
P.~Schweitzer, M.~Strikman, and C.~Weiss, {\it {Intrinsic transverse momentum
  and parton correlations from dynamical chiral symmetry breaking}},  {\em
  JHEP} {\bf 1301} (2013) 163, [\href{http://xxx.lanl.gov/abs/1210.1267}{{\tt
  arXiv:1210.1267}}].

\bibitem{Brodsky:2012ku}
S.~J. Brodsky, C.~D. Roberts, R.~Shrock, and P.~C. Tandy, {\it {Confinement
  contains condensates}},  {\em Phys.Rev.} {\bf C85} (2012) 065202,
  [\href{http://xxx.lanl.gov/abs/1202.2376}{{\tt arXiv:1202.2376}}].

\bibitem{GellMann:1968rz}
M.~Gell-Mann, R.~Oakes, and B.~Renner, {\it {Behavior of current divergences
  under SU(3) x SU(3)}},  {\em Phys.Rev.} {\bf 175} (1968) 2195--2199.

\bibitem{Hornbostel:1990ya}
K.~Hornbostel, {\it {Constructing hadrons on the light cone}},
  \href{http://xxx.lanl.gov/abs/CLNS-90-1038, C90-09-20.1}{{\tt CLNS-90-1038,
  C90-09-20.1}}.

\bibitem{Perry:1994kp}
R.~J. Perry, {\it {Hamiltonian light front field theory and quantum
  chromodynamics}},  \href{http://xxx.lanl.gov/abs/hep-th/9407056}{{\tt
  hep-th/9407056}}.

\bibitem{Zhang:1994ti}
W.-M. Zhang, {\it {Light front dynamics and light front QCD}},  {\em
  Chin.J.Phys.} {\bf 32} (1994) 717--808,
  [\href{http://xxx.lanl.gov/abs/hep-ph/9412244}{{\tt hep-ph/9412244}}].

\bibitem{Burkardt:1995ct}
M.~Burkardt, {\it {Light front quantization}},  {\em Adv.Nucl.Phys.} {\bf 23}
  (1996) 1--74, [\href{http://xxx.lanl.gov/abs/hep-ph/9505259}{{\tt
  hep-ph/9505259}}].

\bibitem{Harindranath:1996hq}
A.~Harindranath, {\it {An Introduction to light front dynamics for
  pedestrians}},  \href{http://xxx.lanl.gov/abs/hep-ph/9612244}{{\tt
  hep-ph/9612244}}.

\bibitem{Brodsky:1997de}
S.~J. Brodsky, H.-C. Pauli, and S.~S. Pinsky, {\it {Quantum chromodynamics and
  other field theories on the light cone}},  {\em Phys.Rept.} {\bf 301} (1998)
  299--486, [\href{http://xxx.lanl.gov/abs/hep-ph/9705477}{{\tt
  hep-ph/9705477}}].

\bibitem{Perry:1997uv}
R.~J. Perry, {\it {Light front QCD: A Constituent picture of hadrons}},
  \href{http://xxx.lanl.gov/abs/hep-th/9710175}{{\tt hep-th/9710175}}.

\bibitem{Miller:1997cr}
G.~A. Miller, {\it {Light front treatment of nuclei: Formalism and simple
  applications}},  {\em Phys.Rev.} {\bf C56} (1997) 2789,
  [\href{http://xxx.lanl.gov/abs/nucl-th/9706028}{{\tt nucl-th/9706028}}].

\bibitem{Venugopalan:1998zd}
R.~Venugopalan, {\it {Introduction to light cone field theory and high-energy
  scattering}},  \href{http://xxx.lanl.gov/abs/nucl-th/9808023}{{\tt
  nucl-th/9808023}}.

\bibitem{Heinzl:2000ht}
T.~Heinzl, {\it {Light cone quantization: Foundations and applications}},  {\em
  Lect.Notes Phys.} {\bf 572} (2001) 55--142,
  [\href{http://xxx.lanl.gov/abs/hep-th/0008096}{{\tt hep-th/0008096}}].

\bibitem{Miller:2000kv}
G.~A. Miller, {\it {Light front quantization: A Technique for relativistic and
  realistic nuclear physics}},  {\em Prog.Part.Nucl.Phys.} {\bf 45} (2000)
  83--155, [\href{http://xxx.lanl.gov/abs/nucl-th/0002059}{{\tt
  nucl-th/0002059}}].

\bibitem{Diehl:2003ny}
M.~Diehl, {\it {Generalized parton distributions}},  {\em Phys.Rept.} {\bf 388}
  (2003) 41--277, [\href{http://xxx.lanl.gov/abs/hep-ph/0307382}{{\tt
  hep-ph/0307382}}].

\bibitem{Belitsky:2005qn}
A.~Belitsky and A.~Radyushkin, {\it {Unraveling hadron structure with
  generalized parton distributions}},  {\em Phys.Rept.} {\bf 418} (2005)
  1--387, [\href{http://xxx.lanl.gov/abs/hep-ph/0504030}{{\tt
  hep-ph/0504030}}].

\bibitem{Mustaki:1994mf}
D.~Mustaki, {\it {Chiral symmetry and the constituent quark model: A null plane
  point of view}},  \href{http://xxx.lanl.gov/abs/hep-ph/9404206}{{\tt
  hep-ph/9404206}}.

\bibitem{Ballesteros:1995mi}
A.~Ballesteros, F.~Herranz, M.~del Olmo, and M.~Santander, {\it {A New 'null
  plane' quantum Poincare algebra}},  {\em Phys.Lett.} {\bf B351} (1995)
  137--145.

\bibitem{Leutwyler:1977vz}
H.~Leutwyler and J.~Stern, {\it {Covariant Quantum Mechanics on a Null Plane}},
   {\em Phys.Lett.} {\bf B69} (1977) 207.

\bibitem{Weinberg:1995mt}
S.~Weinberg, {\it {The Quantum theory of fields. Vol. 1: Foundations}}, .

\bibitem{Weinberg:1996kr}
S.~Weinberg, {\it {The quantum theory of fields. Vol. 2: Modern applications}},
  .

\bibitem{Weinberg:1995fe}
S.~Weinberg, {\it {Unbreaking symmetries}},  {\em Conf.Proc.} {\bf C930308}
  (1993) 3--11.

\bibitem{'tHooft:1973jz}
G.~'t~Hooft, {\it {A Planar Diagram Theory for Strong Interactions}},  {\em
  Nucl.Phys.} {\bf B72} (1974) 461.

\bibitem{Adler:1965ka}
S.~L. Adler, {\it {Calculation of the axial vector coupling constant
  renormalization in beta decay}},  {\em Phys.Rev.Lett.} {\bf 14} (1965)
  1051--1055.

\bibitem{Weisberger:1965hp}
W.~I. Weisberger, {\it {Renormalization of the Weak Axial Vector Coupling
  Constant}},  {\em Phys.Rev.Lett.} {\bf 14} (1965) 1047--1051.

\bibitem{Gilman:1967qs}
F.~J. Gilman and H.~Harari, {\it {Strong Interaction Sum Rules for Pion-Hadron
  Scattering}},  {\em Phys.Rev.} {\bf 165} (1968) 1803--1829.

\bibitem{Beane:1999hp}
S.~R. Beane, {\it {Low-energy constants from high-energy theorems}},  {\em
  Phys.Rev.} {\bf D61} (2000) 116005,
  [\href{http://xxx.lanl.gov/abs/hep-ph/9910525}{{\tt hep-ph/9910525}}].

\bibitem{Beane:2002ud}
S.~R. Beane and U.~van Kolck, {\it {The Role of the Roper in QCD}},  {\em
  J.Phys.} {\bf G31} (2005) 921--934,
  [\href{http://xxx.lanl.gov/abs/nucl-th/0212039}{{\tt nucl-th/0212039}}].

\bibitem{Beane:2002td}
S.~R. Beane and M.~J. Savage, {\it {A Conjecture about hadrons}},  {\em
  Phys.Lett.} {\bf B556} (2003) 142--152,
  [\href{http://xxx.lanl.gov/abs/hep-ph/0212106}{{\tt hep-ph/0212106}}].

\bibitem{Weinberg:1978kz}
S.~Weinberg, {\it {Phenomenological Lagrangians}},  {\em Physica} {\bf A96}
  (1979) 327.

\bibitem{Gasser:1983yg}
J.~Gasser and H.~Leutwyler, {\it {Chiral Perturbation Theory to One Loop}},
  {\em Annals Phys.} {\bf 158} (1984) 142.

\bibitem{Bardeen:1979xx}
W.~A. Bardeen, R.~B. Pearson, and E.~Rabinovici, {\it {Hadron Masses in Quantum
  Chromodynamics on the Transverse Lattice}},  {\em Phys.Rev.} {\bf D21} (1980)
  1037.

\bibitem{Burkardt:1998ws}
M.~Burkardt and H.~El-Khozondar, {\it {Light front QCD with Wilson fermions on
  a transverse lattice}},  {\em Phys.Rev.} {\bf D60} (1999) 054504,
  [\href{http://xxx.lanl.gov/abs/hep-ph/9805495}{{\tt hep-ph/9805495}}].

\bibitem{Dalley:1998bj}
S.~Dalley and B.~van~de Sande, {\it {Transverse lattice approach to light front
  Hamiltonian QCD}},  {\em Phys.Rev.} {\bf D59} (1999) 065008,
  [\href{http://xxx.lanl.gov/abs/hep-th/9806231}{{\tt hep-th/9806231}}].

\bibitem{Dalley:2001gj}
S.~Dalley, {\it {Mesons on a transverse lattice}},  {\em Phys.Rev.} {\bf D64}
  (2001) 036006, [\href{http://xxx.lanl.gov/abs/hep-ph/0101318}{{\tt
  hep-ph/0101318}}].

\bibitem{Burkardt:2001mf}
M.~Burkardt and S.~Seal, {\it {A Study of light mesons on the transverse
  lattice}},  {\em Phys.Rev.} {\bf D65} (2002) 034501,
  [\href{http://xxx.lanl.gov/abs/hep-ph/0102245}{{\tt hep-ph/0102245}}].

\bibitem{Burkardt:2001dy}
M.~Burkardt and S.~K. Seal, {\it {A Study of heavy light mesons on the
  transverse lattice}},  {\em Phys.Rev.} {\bf D64} (2001) 111501,
  [\href{http://xxx.lanl.gov/abs/hep-ph/0105109}{{\tt hep-ph/0105109}}].

\bibitem{Burkardt:2001jg}
M.~Burkardt and S.~Dalley, {\it {The Relativistic bound state problem in QCD:
  Transverse lattice methods}},  {\em Prog.Part.Nucl.Phys.} {\bf 48} (2002)
  317--362, [\href{http://xxx.lanl.gov/abs/hep-ph/0112007}{{\tt
  hep-ph/0112007}}].

\bibitem{Dalley:2002nj}
S.~Dalley and B.~van~de Sande, {\it {Transverse lattice calculation of the pion
  light cone wavefunctions}},  {\em Phys.Rev.} {\bf D67} (2003) 114507,
  [\href{http://xxx.lanl.gov/abs/hep-ph/0212086}{{\tt hep-ph/0212086}}].

\bibitem{Bratt:2004wq}
J.~Bratt, S.~Dalley, B.~van~de Sande, and E.~Watson, {\it {Small-x behaviour of
  lightcone wavefunctions in transverse lattice gauge theory}},  {\em
  Phys.Rev.} {\bf D70} (2004) 114502,
  [\href{http://xxx.lanl.gov/abs/hep-ph/0410188}{{\tt hep-ph/0410188}}].

\bibitem{Banks:1979yr}
T.~Banks and A.~Casher, {\it {Chiral Symmetry Breaking in Confining Theories}},
   {\em Nucl.Phys.} {\bf B169} (1980) 103.

\bibitem{Witten:1979kh}
E.~Witten, {\it {Baryons in the 1/n Expansion}},  {\em Nucl.Phys.} {\bf B160}
  (1979) 57.

\bibitem{Beane:1998xp}
S.~Beane, {\it {Chiral multiplets of large N(c) ground state baryons}},  {\em
  Phys.Rev.} {\bf D59} (1999) 031901,
  [\href{http://xxx.lanl.gov/abs/hep-ph/9809328}{{\tt hep-ph/9809328}}].

\bibitem{Gervais:1983wq}
J.-L. Gervais and B.~Sakita, {\it {Large N QCD Baryon Dynamics: Exact Results
  from Its Relation to the Static Strong Coupling Theory}},  {\em
  Phys.Rev.Lett.} {\bf 52} (1984) 87.

\bibitem{Dashen:1993as}
R.~F. Dashen and A.~V. Manohar, {\it {Baryon - pion couplings from large N(c)
  QCD}},  {\em Phys.Lett.} {\bf B315} (1993) 425--430,
  [\href{http://xxx.lanl.gov/abs/hep-ph/9307241}{{\tt hep-ph/9307241}}].

\bibitem{Dashen:1993jt}
R.~F. Dashen, E.~E. Jenkins, and A.~V. Manohar, {\it {The 1/N(c) expansion for
  baryons}},  {\em Phys.Rev.} {\bf D49} (1994) 4713,
  [\href{http://xxx.lanl.gov/abs/hep-ph/9310379}{{\tt hep-ph/9310379}}].

\end{thebibliography}\endgroup

\end{document}